\documentclass[sigconf]{acmart}

\newcommand{\hide}[1]{}
\newcommand{\eat}[1]{}

\usepackage{booktabs} % For formal tables

\renewcommand\footnotetextcopyrightpermission[1]{} % removes footnote with conference information in first column
\usepackage{hyperref}
\usepackage{balance}

\usepackage{color}
\usepackage{multirow}
\usepackage{subfigure}
\usepackage{enumerate}

\usepackage{algorithm}
\usepackage{algpseudocode}
\usepackage{xspace}
\newtheorem{definition}{Definition}

\newtheorem{lemma}{Lemma}

\newcommand{\fix}[1]{[\textcolor{blue}{#1}]}

\newcommand{\panos}[1]{[\textcolor{magenta}{Panos: #1}]}
\newcommand{\stitle}[1]{\vspace{1ex}\noindent\textbf{#1}}

\newcommand{\onelevel}{\ensuremath{\mathsf{1\mbox{-}level}}\xspace}
\newcommand{\twolevel}{\ensuremath{\mathsf{2\mbox{-}level}}\xspace}
\newcommand{\twolevelplus}{\ensuremath{\mathsf{2\mbox{-}level^+}}\xspace}
\newcommand{\refvone}{\textsf{Simple}\xspace}
\newcommand{\refvtwo}{\textsf{RefAvoid}\xspace}
\newcommand{\refvthree}{\ensuremath{\mathsf{RefAvoid^+}}\xspace}

\newcommand{\qatomic}{\ensuremath{\mathsf{queries}\mbox{-}\mathsf{based}}\xspace}
\newcommand{\tatomic}{\ensuremath{\mathsf{tiles}\mbox{-}\mathsf{based}}\xspace}

\pdfoutput=1

\begin{document}

\title{A Two-level Spatial In-Memory Index*}\thanks{* Preliminary version of the IEEE ICDE'21 paper titled ``A Two-layer Partitioning for Non-point Spatial Data'' }

\author{Dimitrios Tsitsigkos}
\affiliation{%
  \institution{University of Ioannina, Greece}
  \institution{Athena Research Center, Greece}
}
\email{dtsitsigkos@cse.uoi.gr}
%\email{dtsitsigkos@imis.athena-innovation.gr}

\author{Konstantinos Lampropoulos}
\affiliation{%
  \institution{University of Ioannina, Greece}
}
\email{klampropoulos@cse.uoi.gr}

\author{Panagiotis Bouros}
\orcid{0000-0002-8846-4330}
\affiliation{%
  \institution{Johannes Gutenberg University Mainz, Germany}
}
\email{bouros@uni-mainz.de}

\author{Nikos Mamoulis}
\affiliation{%
  \institution{University of Ioannina, Greece}
}
\email{nikos@cse.uoi.gr}

\author{Manolis Terrovitis}
\affiliation{%
  \institution{Athena Research Center, Greece}
}
\email{mter@imis.athena-innovation.gr}

\begin{abstract}
Very large volumes of spatial data increasingly become available and
demand effective management.
While there has been decades of research on spatial data management,
few works consider the current state of commodity hardware, having
relatively large memory and the ability of parallel multi-core
processing.
In this paper, we re-consider the design of spatial indexing under
this new reality.
Specifically, we propose a main-memory indexing approach
for objects with spatial extent, which
is based on a classic regular space partitioning into disjoint tiles.
The novelty of our index is that the contents of each tile are further
partitioned into four classes.
This second-level partitioning not only
reduces the number of comparisons required to compute the results,
but also avoids the generation and elimination of duplicate results,
which is an inherent problem of spatial indexes based on disjoint
space partitioning.
The spatial partitions defined by our indexing scheme are totally
independent, 
facilitating effortless parallel evaluation, as no
synchronization or communication between the partitions is necessary.
We show how our index can be used to efficiently process spatial range
queries and drastically reduce the cost of the refinement step of the
queries.
In addition, we study the efficient processing of numerous range
queries in batch and in parallel.
Extensive experiments on real datasets confirm the efficiency of our approaches.
\end{abstract}

%\keywords{}

\maketitle

\section{Introduction}
The management and indexing of spatial data
has been studied
extensively for at least four decades.
Classic spatial indexes \cite{GaedeG98} have been
designed for the -- now obsolete -- storage model of the 80's, i.e.,
the data are too big to reside in memory and the goal is to minimize
the I/O cost during query evaluation. Things have changed a lot since
then. First, memories have become much bigger and cheaper.
In most applications, the spatial data can easily fit in the memory of
even a commodity machine. Second, modern processors have multiple
cores and facilitate parallel query processing.
In this paper, we re-consider the design of spatial indexing under
this new reality. Our goal is a main-memory spatial index,
which outperforms the state-of-the-art spatial access methods,
considering computational cost as the main factor.

Our index is based on a simple grid-based space partitioning.
Grid-based indexing
%is not sophisticated and
%which
has several advantages over
hierarchical indexes, such as the R-tree \cite{Guttman84}.
First, the
relevant partitions to a query are very fast to identify (using
algebraic operations only).
Second, the partitions are
totally independent to each other and can be handled by different
threads without the need of any synchronization or scheduling.
Third, updates can be performed very fast, as locating
the cell
which contains or the cells which intersect an object
takes constant time
and no changes to the space partitioning are required.
Hence, main memory grids have been preferred over hierarchical
indexes,
especially for the (in-memory) management of highly dynamic collections
of 2D points \cite{MokbelXA04,KalashnikovPH04,YuPK05,PapadiasMH05,SidlauskasSCJS09,RayBG14}.

\begin{figure}[htb]
  \centering
  \subfigure[grid]{
    \label{fig:grid}
     \includegraphics[width=0.37\columnwidth]{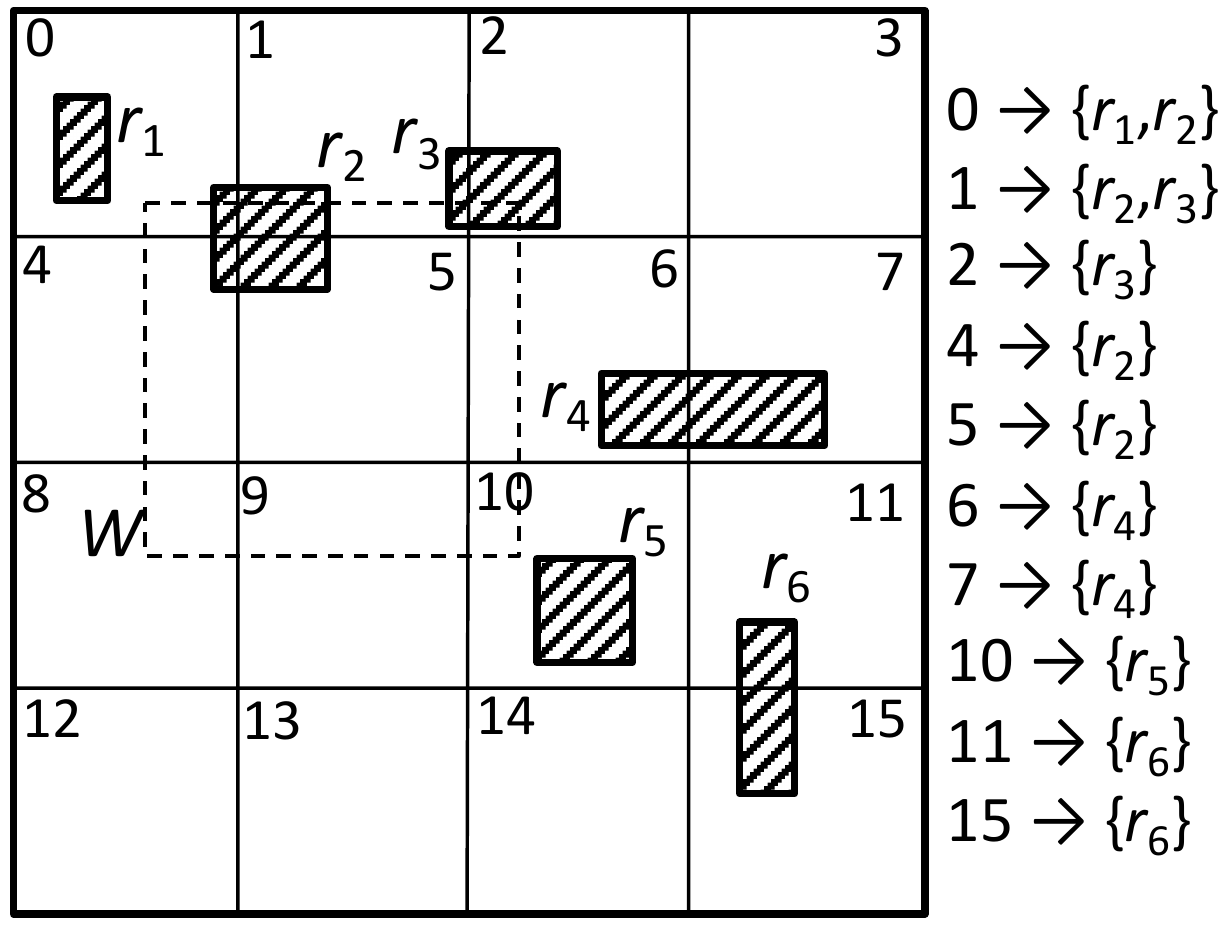}
    }
   \hspace{0in}
  \subfigure[R-tree]{
    \label{fig:rtree}
    \includegraphics[width=0.55\columnwidth]{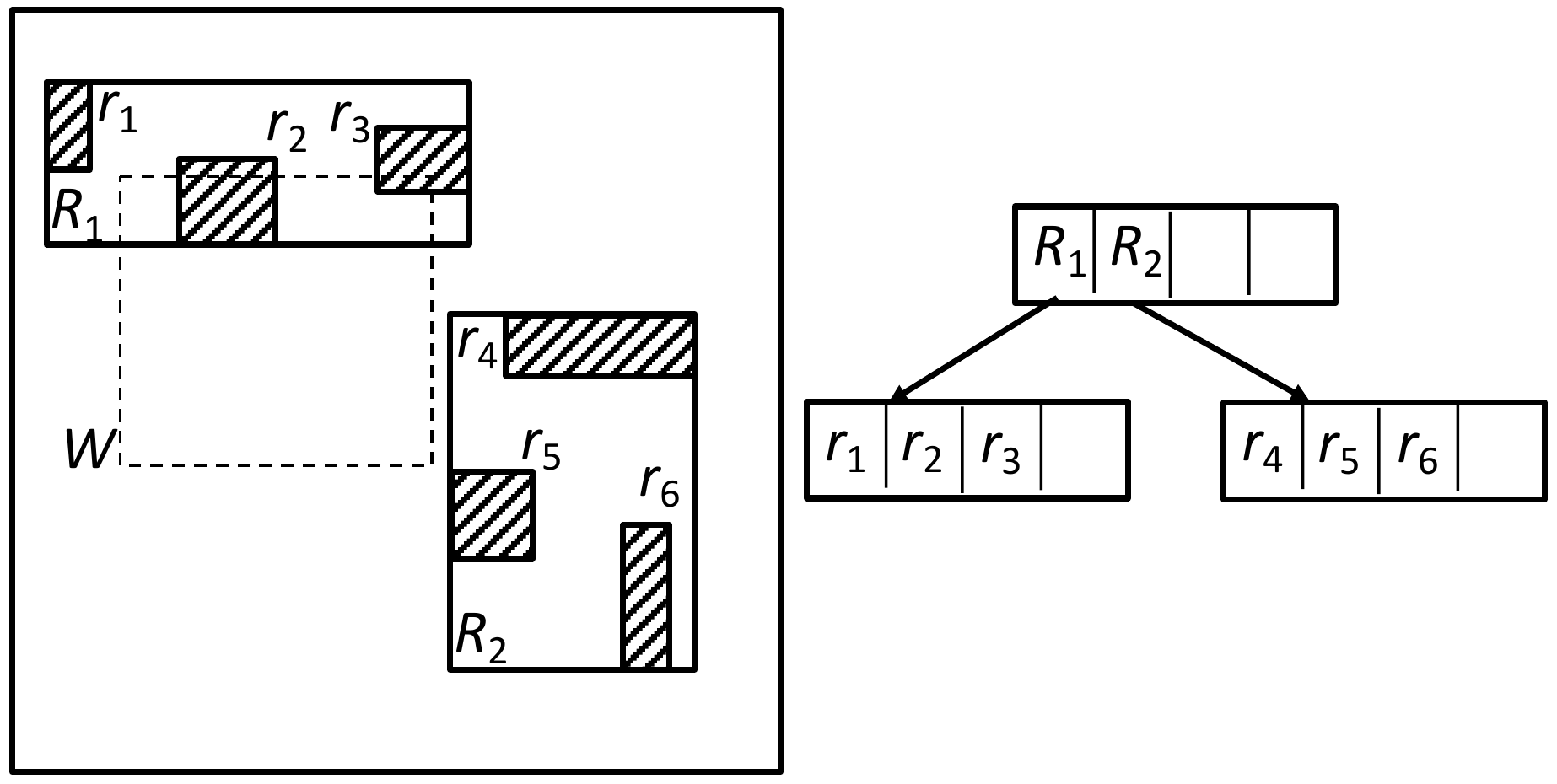}
  }
  \vspace{-0.1in}
  \caption{Examples of indexing schemes}\label{fig:intro}
\end{figure}

Still, spatial grids have their own
weaknesses. First, the distribution of the objects to cells can
be highly uneven, which renders some of the partitions to be
overloaded.
This issue can be alleviated by increasing the grid
granularity, which, however, may result in numerous empty tiles.
The problem of empty tiles can be easily handled by using a
hash table or a bitmap to mark non-empty tiles, or by modeling and
searching non-empty (fine) tiles using space-filling curves \cite{JensenLO04,DittrichBS09}.  
The most important issue arises in the indexing of non-point objects
(e.g., polygons), which are typically approximated by their {\em minimum
bounding rectangles} (MBRs).
If an MBR 
intersects multiple tiles, then it is assigned to multiple partitions,
which causes replication, increases query times, and requires
special handing for possible duplicate results.
For example, consider the six object MBRs depicted in Figure
\ref{fig:intro}(a), partitioned using a 4$\times$4 grid.
Some MBRs (e.g., $r_2$) are assigned to multiple tiles. Besides the
increased space requirements due to replication, given a query range
(e.g., $W$), a replicated object (e.g., $r_2$) may be examined and
reported multiple times (e.g., at tiles 0, 1, 4, and 5).
Spatial indexes that allow overlapping partitions (such as the R-tree)
do not have this problem because each object falls into exactly one
leaf node, so there is no replication and no need for duplicate result
avoidance. For example the R-tree of Figure
\ref{fig:grid} would examine the MBR of $r_2$ only once, in the
leaf node under entry $R_1$.
Still, as mentioned above, R-tree like methods have
relatively high update costs due to index maintenance 
and query evaluation using them is harder
to parallelize.

Our goal is the design of a grid based index for non-point data
which maintains the
simplicity and advantages of a grid, without having its disadvantages.
The niece of our index is the introduction an additional,
secondary-level partitioning which divides the objects that intersect
a tile into four classes ($2^m$ classes in an $m$-dimensional space),
based on whether their MBRs start inside or before the tile in each
dimension.
This division helps us to avoid accessing some classes of objects
during query evaluation, which reduces the query cost
and at the same time avoids duplicate results, while ensuring that
no results are missed. For example, in Figure \ref{fig:rtree}, due
to our second-level partitioning object $r_2$ will be examined and
reported only by tile 0 because both the object starts in
that tile and the query starts in or before that tile.
We lay out the set of rules, based on which range queries using our
index are evaluated.
Besides, we show how our index can be used to answer distance range
queries, reducing the number of expensive distance computations that
alternative indexing schemes require.
Finally, we show how
%, via our index,
the number of applications of the expensive
refinement step can greatly be reduced.

Besides rectangular range queries, we also study the evaluation of
circular range (i.e., disk) queries and, in general, queries with
convex range shapes. We show how our indexing scheme reduces the
number of comparisons and avoids duplicate results, also in this
case. In addition, we show how for the majority of query results, the
refinement step can be avoided by a simple post-filtering test on the
object MBRs. Finally, we investigate the efficient evaluation of
numerous range queries in batch and in parallel.

We compare our index experimentally with a state-of-the-art
implementation of an in-memory R-tree from boost.org and show that it
is up to several times faster, especially for large queries on large
datasets.
In addition, our index performs much better that the R-tree for mixed
workloads (with inserts and range queries).
We also show that our approach (which is directly parallelizable)
scales gracefully with the number of
cores (i.e., threads in a multi-core machine),
making it especially suitable for shared-nothing parallel environments where
tree-based spatial indexes 
(such as the R-tree) are hard to deploy.

%\fix{add stuff on parallelism}

%Our index turns out to be faster than the R-tree for range queries and
%updates \fix{++} 

In summary, this paper makes the following contributions:
\begin{itemize}
\item
We design a novel secondary-level partitioning approach for
space-partitioning indexes (such as grids). 
\item
We show how spatial range queries can benefit from our indexing
scheme, by avoiding redundant comparisons and the generation of
duplicate results.
\item
We introduce a simple additional filter that avoids the refinement
step for the great majority of the objects, rendering range queries
very efficient in practice.
\item
We conduct an extensive experimental evaluation which demonstrates
the superiority of our index in comparison to alternative methods and
its scalability when evaluating multiple queries using multiple cores. 
\end{itemize}

% TODO
% disk query: fix problem with duplicates
% disk query: R-tree implementation
% boost R-tree: construct and check packed R-trees
% range queries: find additional competitors, if possible
% parallelism: implement and experiment with four alternatives:
% (a) independent jobs,
% (b) evaluation of many range queries as a parallel spatial join
% (c) query breakdown to multiple mini-tasks
% (d) scheduling per tile for a query batch
% index design: cost model for best grid granularity
% updates: test mixed workloads of queries and updates
% refinement step: test avoidance, simple test and sophisticated(?)
% quadtree: (a) as competitor (b) application of our idea to quadtree

The rest of the paper is organized  as follows.
Section \ref{sec:related} provides the necessary background and
discusses related work.
Section \ref{sec:index} introduces our secondary partitioning scheme
and its application in grid-based spatial indexing.
Section \ref{sec:range} shows how spatial range query evaluation can
benefit from our indexing scheme.
In Section \ref{sec:refine}, we present a filtering condition that
applies on the MBRs of the objects and can be used to confirm the
inclusion of an object to a range query result, without the need of a
refinement step.
Section \ref{sec:parallel} discusses how numerous range queries that
may need to be handled can be
processed efficiently and in parallel.
An experimental evaluation is presented in Section \ref{sec:exps}.
Finally, Section \ref{sec:conclusion} concludes the paper with a
discussion about future work.

\section{Background and Related Work}
\label{sec:related}
In this section, we introduce the necessary background in spatial data management \cite{2011Mamoulis} and present related work to our research.
%Specifically, we discuss basic spatial data and query types
%and principles for query evaluation.\fix{++}

%\fix{define spatial data and queries; review the filter-refinement
%  framework; review classic spatial indexing prototypes; review recent
%  work that considers modern hardware} 

\subsection{Spatial Data and Queries}
\label{sec:prelims:queries}
Common types of spatial objects include points (defined by one value per dimension), rectangles (defined by one interval per dimension), line segments (defined by a pair of points), polygons (defined by a sequence of points), linestrings (defined by a sequence of points), etc. 
%Figure 1(a) shows examples of these datatypes on the plane. 
%To characterize the relative position between two spatial objects,
Three classes of spatial relationships
characterize the relative position and geometry of spatial objects. 
\emph{Topological} relationships model the relation between the geometric extents of objects (e.g., overlap, inside).
\emph{Directional} relationships compare the relative locations of the objects with respect to a coordinate (or cardinal) system (e.g., north/south, above/below).
Last, \emph{distance} relationships capture distance/proximity information between two objects (e.g., near/far). 
%Figure 1(b) illustrates examples of spatial relationships.
%In this work, our focus is on query predicates defined with topological relationships for spatial objects with extent, e.g., rectangles and polygons; directional and distance-based predicates can also be easily handled and are left as future work.

The most frequently applied query operation on spatial data is the spatial \emph{selection} or \emph{range query} which retrieves the objects that
satisfy a spatial relationship with a reference spatial object. Typically, the reference object is a region $W$ and the objective is to retrieve the objects that intersect $W$ or are inside $W$. In another popular range query type (especially, in location-based services), given a reference location $q$ and a distance threshold $\epsilon$, the objective is to find all objects having at most $\epsilon$ Euclidean distance from $q$.
%for example intersect or are inside a well-defined area\eat{, or are within some distance from a reference object}. Formally:
Spatial access methods are primarily designed for spatial selection queries. Other important spatial queries include nearest neighbor queries, intersection joins and distance joins.

\subsection{Query Processing Principles}
\label{sec:prelims:qprocess}
The potentially complex geometry of the objects renders inefficient the evaluation of spatial predicates directly on their exact representation.
Hence, spatial queries are processed in two steps following a \emph{filtering-and-refinement} framework.
% \eat{, illustrated in Figure~\ref{fig:filter-refine}}.
During the \emph{filtering} step, the query is applied on
the \emph{Minimum Bounding Rectangles} (MBRs),
which approximate the objects.
If the MBR of an object
%(or the MBRs, in case of a spatial join)
does not qualify the query predicate, then the exact geometry does not qualify it either.
%This can be demonstrated in Figure~\ref{fig:filter-refine}(b) where MBRs that do not intersect $W$ and so, do not enclose objects that qualify the query either.
The filtering step is a computationally cheap way to prune the search space, in many cases powered by spatial indexing, but it only provides candidate query results. During the \emph{refinement} step, the exact representations of the candidates are tested with the query predicate to retrieve the actual results.
%In Figure~\ref{fig:filter-refine}(c), observe that only two from the three candidates are query results.
%In some cases, the refinement step can be avoided. For example, if at least one side of the MBR is inside the query range $W$, then the object definitely intersects $W$.

%spatial query evaluation is typically performed in two steps, also know as the \emph{filter-and-refinement} framework. In the filter step, the query is applied on the minimun bounding rectangles (MBRs) of the objects, which are simple lightweight approximations; see for example the MBR of a polygon in \todo{Figure~\ref{}}. Object MBRs (or pairs of object MBRs) that do not qualify the query can be pruned in the filter step, while for the objects (or object pairs) that pass the filter an expensive refinement step is applied using their exact geometries. For example, if the MBR of a river does not intersect the MBR of a city, there is no chance that the exact geometries of the two objects intersect.

% \subsection{Partitioning and Indexing}
% \label{sec:prelims:index}

%\section{Related Work}
%\label{sec:related}
%\fix{Related work goes here (or before the conclusion)}
%In this section, we review the most related literature on spatial data management, discussing the key issues of partitioning and indexing spatial objects. We also review approaches and systems for parallel and distributed management of spatial data.

\subsection{Spatial Partitioning and Indexing}
\eat{
\begin{figure}[t]
\centering
  \includegraphics[width=0.99\columnwidth]{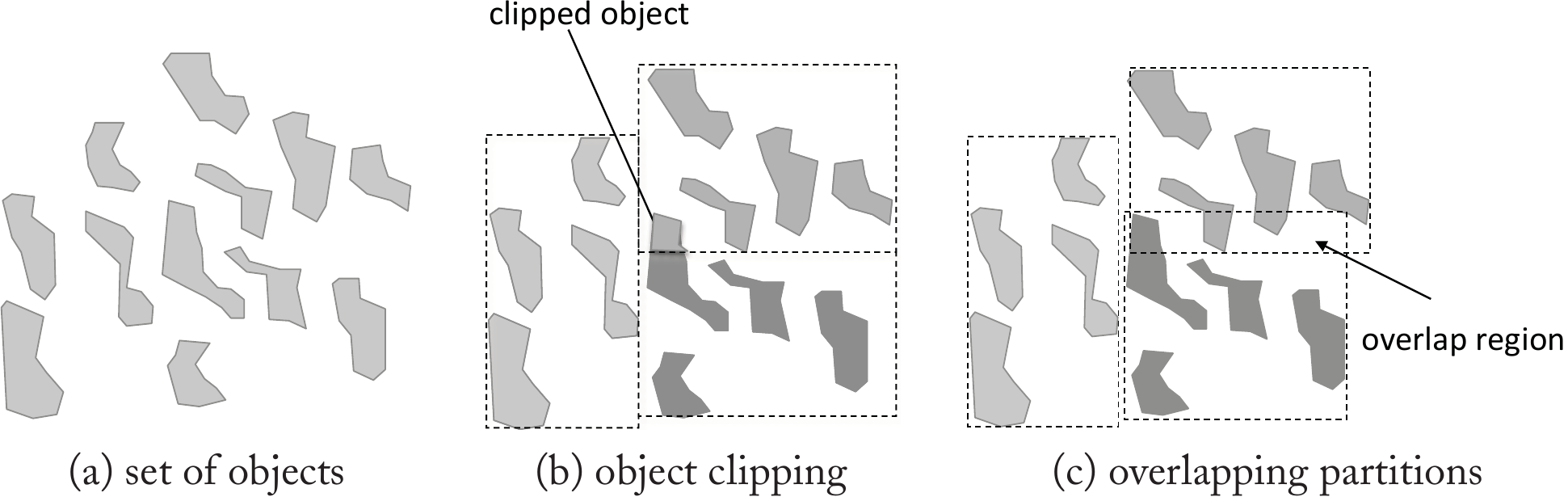}
  \caption{Object clipping on space-oriented partitioning vs. overlapping partitions on data-oriented partitioning (picture from \cite{2011Mamoulis}).}
  \label{fig:partitioning}
\end{figure}
}
Due to the complex nature of spatial objects, decades of research efforts have been devoted on spatial indexing; as a result, a number of \emph{Spatial Access Methods} (SAMs) have been proposed\eat{. A detailed discussion of spatial indexing can be found in} \cite{Samet90,Samet06}. Although the majority of SAMs focus on disk-resident data, it is straightforward to also use them in main memory, which is the focus of our work. Typically, the goal of an SAM is to group closely located objects in space, into the same index blocks (traditionally, blocks are disk pages). These blocks are then organized in an index (single-level or  hierarchical).

Depending on the nature of the partitioning, spatial indices can be classified into two classes \cite{OlmaTHA17}.
%Early works in spatial indexing primarily focused on point data. The simplest indexing structure is a grid which uniformly divides the object space into cells (partitions), using axis-parallel hyperplanes. 
Indices based on \emph{space-oriented} partitioning divide the space into disjoint partitions\eat{, independent of the data distribution}. As a result, objects whose extent overlaps with multiple partitions need to be replicated (or {\em clipped}) in each of them.
%; this technique
% , illustrated in Figure~\ref{fig:partitioning}(b),
% is known as object \emph{clipping}.
A grid \cite{BentleyF79} is the simplest index based on space-oriented partitioning; the space is uniformly divided into cells (partitions), using axis-parallel lines. Hierarchical indices that fall in this category are the
kd-tree \cite{Bentley75} and the quad-tree \cite{FinkelB74}.
Space-oriented partitioning was originally proposed and is especially suitable for
indexing collections of points,
because no replication issues arise.
%Recent indexes for point data 
A bitmap-based index for point data was recently proposed in \cite{NagarkarCB15}.
SIDI \cite{NguyenDP16} is another spatial index for point data, which
learns the characteristics of the dataset before construction and
its layout is designed to fit the data well.

For non-point objects,
the replication of object MBRs to multiple space-oriented partitions
may negatively affect query performance.
%\emph{space-oriented} partitioning
%If excessive,
%replication of non-point objects to multiple partitions
%may affect query performance.
In addition, due to object replication, the same query results may be detected in
multiple partitions and deduplication techniques should be applied, as discussed in the Introduction.
% of the index.
In view of this, an alternative class
of indices, based on a \emph{data-oriented} partitioning, were also
proposed, allowing the extents of the partitions to overlap and
ensuring that their contents are disjoint
(i.e., each object is assigned
to exactly one partition).
%(i.e., no replication).
%way to index spatial objects is to allow non-disjoint or
%overlapping partitions.
% ; Figure~\ref{fig:partitioning}(c) illustrates an example.
%and approximately the same 
%with approximately the same number of contents.
The R-tree \cite{Guttman84} (and its variants, e.g., the R*-tree \cite{BeckmannKSS90})
is the most popular SAM in this class (and in general).
The R-tree is a height-balanced tree, which generalizes the B$^+$-tree in the multi-dimensional space and hierarchically groups object MBRs to blocks.
Each block is also approximated by an MBR, hence the tree defines a hierarchy of MBR groups.
Some R-tree variants use circles (or spheres in the 3D space) instead of MBRs, i.e., the SS-tree \cite{WhiteJ96}, or a combination of circles and rectangles, i.e., the SR-tree \cite{KatayamaS97}.

Most spatial indexing methods are designed for the efficient evaluation of spatial range queries.
% Without loss of generality, consider an overlap predicate and a rectangular query region $W$.
In brief, during the filtering step, the goal is to determine which partitions of the space intersect the query region $W$.
In case of hierarchical indices such as the kd-tree, the quad-tree and the R-tree, the query is processed by recursively traversing the nodes whose MBRs intersect $W$, starting from the root.
% In case of space-filling curves, the query region is modeled as a set of intervals in the 1D space of the curve values, and the spatial range query is transformed to a set of range queries on a relational index, e.g., a B$^+$-tree.
Finally, every object whose MBR overlaps with region $W$ is passed as a candidate to the refinement step where its exact geometry is compared against $W$\eat{ to output the final results}.

Most spatial indexes have been designed to support dynamic updates. The R*-tree \cite{BeckmannKSS90}  differs from the original  R-tree \cite{Guttman84} in its insertion algorithm, which is designed to be both efficient and to result in a high tree quality.
Bulk loading methods for R-trees have also been proposed, with the most popular method being the sort-tile-recursive approach \cite{LeuteneggerEL97}. Naturally, updates on hierarchical indexes are more expensive, compared to updates on single-level (flat) indexes, because they may result in index reorganization. Hence, single-level indexes, such as grids are preferred over hierarchical ones in workloads with many updates (e.g., when indexing moving objects \cite{KalashnikovPH04}).

%\subsection{Main memory SAMs}
% ,  is the most popular SAM. It is a height-balanced tree that hierarchically indexes the MBRs of the spatial objects.
The R-tree was originally proposed for disk-resident data with the key focus on minimizing the I/O during query processing. The CR-tree \cite{KimCK01} is an optimized R-tree for the memory hierarchy. 
BLOCK \cite{OlmaTHA17}
is a recently proposed main-memory spatial
index, which uses a hierarchy of grids.
At each
level, a uniform grid with higher resolution compared to the level above is used.
Given a range query, starting from the uppermost grid, BLOCK evaluates the query on cells that are completely contained in the query. The remaining query parts (excluding the cells that are contained in the range) are either evaluated at the cells they overlap at the current level, or they are evaluated recursively at the level below, depending on the estimated benefit.

\subsection{Parallel and Distributed Data Management}

Early efforts on parallel and distributed spatial query evaluation have mainly focused on spatial joins, which are more expensive than range queries and they can benefit more from parallelism.  
%\cite{RaySBJ14}
%\cite{BrinkhoffKS96}
The R-tree join (RJ) algorithm \cite{BrinkhoffKS93} and PBSM \cite{PatelD96} were
parallelized in \cite{BrinkhoffKS96} and \cite{ZhouAT97,PatelD00,RaySBJ14}, respectively.
%Intersecting pairs of entries in two tested R-tree nodes by RJ define independent join tasks for the corresponding R-trees, so they can be parallelized (using a shared-memory model) \cite{BrinkhoffKS96}.
%Similarly, joining pairs of tiles in PBSM can be performed in parallel and/or distributed manner \cite{ZhouAT97,PatelD00}.
%Ray et al. \cite{RaySBJ14} proposed a parallel quad-tree based spatial join algorithm,
%focusing on load balancing and on
%the minimizing the cost of the refinement step.
%Not only the MBRs of the objects, but also their exact geometries
%are clipped and partitioned, in order to apply the refinement step only in parts
%of the geometries that may overlap.
%s (that may be handled by different nodes) in
%order to avoid any communication between nodes when object pairs
%whose MBRs intersect are refined.

With the advent of Hadoop, research on spatial data management has shifted to the development of distributed spatial data management systems \cite{AjiWVLL0S13,EldawyM15,YouZG15,XieL0LZG16,YuZS19}.
{\em Hadoop-GIS} \cite{AjiWVLL0S13} is one of the first efforts in this direction.
Spatial data in Hadoop-GIS are partitioned using a hierarchical grid, wherein high density tiles are split to smaller ones, in order to handle data skew.
The nodes of the cluster share a {\em global tile index} which can be used to find
the HDFS files where the contents of the tiles are stored.
For query evaluation, an implicit parallelization approach is followed, which leverages MapReduce.
That is, the partitioned objects are given IDs based on the tiles they reside and
finding the objects in each tile can be done by a map operation. Spatial queries are
implemented as MapReduce workloads. Duplicate results in spatial queries are eliminated by adding a MapReduce job at the end. 
In the {\em SpatialHadoop} system \cite{EldawyM15} data are also spatially partitioned, but
different options for partitioning based on different spatial indexes are possible (i.e., grid based, R-tree based, quadtree based, etc.)
Different spatial datasets could be partitioned by a different approach.
A global index for each dataset is stored at a Master node, indexing
for each HDFS file block the MBR of its contents.
In addition, a local index is built at each physical partition and used by map tasks.

Spark-based implementations of spatial data management systems
\cite{YouZG15,XieL0LZG16,YuZS19}
apply similar
partitioning approaches. The main difference to Hadoop-based implementations is
that data, indexes, and intermediate results are shared in
the memories of all nodes in the cluster 
as {\em resilient distributed datasets} (RDDs)
and can be made persistent on disk.
Unlike SpatialSpark \cite{YouZG15} and GeoSpark \cite{YuZS19} which are built on top of Spark, Simba \cite{XieL0LZG16} has its own native query engine and query optimizer, however, Simba does not support non-point geometries.
Pandey et al. \cite{PandeyKNK18}
conduct a comparison between in-memory spatial analytics systems and find that they
scale well in general, although each one has its own limitations.
Similar conclusions are drawn in another study \cite{AlamRB18}.

We observe that distributed spatial data management systems focus more on data
partitioning and less on minimizing the cost of query evaluation
at each partition. In other words, emphasis is given on scaling
out (i.e., making the cost anti-proportional to the number of
nodes), rather than on per-node scalability (i.e., reducing the
computational cost per node) and multi-core parallelism.
For example, a typical range query throughput rate reported  by the
tested systems in \cite{PandeyKNK18,AlamRB18} is a few hundred queries per
minute, whereas for the same scale of data
an in-memory R-tree can handle on a single machine (without
parallelism) tens of thousands of queries per minute
(according to our tests in Section \ref{sec:exps}).

\section{Two-level Spatial Partitioning}
\label{sec:index}
We consider the classic approach of approximating spatial objects by
their {\em minimum bounding rectangles} (MBRs).
By imposing a $N\times M$ regular grid over the space, we can divide
it into $N\cdot M$ disjoint {\em tiles}.
Determining $N$ and $M$ is not a subject of this section; we will
discuss/study this issue in Section \ref{sec:exps}.
Each tile divides a {\em spatial partition}. An object $o$ is assigned to a tile $T$
if $MBR(o)$ and $T$ intersect (i.e., they have at least one common
point); in this case, $o$ is assigned to tile $T$.
Since $MBR(o)$ can intersect with multiple tiles, $o$ can be assigned
to more than one tiles. We target applications where the object
extents are relatively small compared to the map (and to the extent of
a tile); hence object
replication is expected to be low. 

For example, Figure \ref{fig:index1} shows a grid and a spatial
object $o_1$,
%(colored darkgrey),
whose MBR intersects tiles $T_a$ and $T_b$; $o_1$ is
assigned to both tiles.
%\fix{add example}
For each tile $T$, we keep a list of (MBR, object-id) pairs
that are assigned to $T$.
For example, the MBR and id of $o_1$
in Figure \ref{fig:index1}
%(paired with a reference to the object)
appears in the list of $T_a$ and $T_b$.  
%\fix{continue example}
This means that while the MBRs and ids of the objects can be replicated
to multiple tiles, the actual geometry of an object is stored only
once in a separate data structure (e.g., an array or a hash-map)
in order to be retrieved fast, given the object's id.
%\fix{to be
%  defined more precisely}.
Since the spatial distribution of objects may not be uniform, there
could be empty tiles.
If the number of empty tiles is significantly large compared to the total
number of tiles, we can use a hash-map to assign each non-empty tile
to the set of rectangles in it.
%\fix{to be  defined more precisely}.
The above storage scheme is quite effective for main-memory data because
it supports queries and updates quite fast, while it is
straightforward to parallelize popular spatial queries and operations.

\subsection{Evaluating queries over a simple grid}
\label{sec:index:grid}
We now discuss in more detail how this simple indexing scheme can be
used to evaluate rectangular range queries and expose its limitations.
We first introduce some notation that will also be useful
when we discuss our solution.

\begin{figure}[t]
\centering
  \includegraphics[width=0.75\columnwidth]{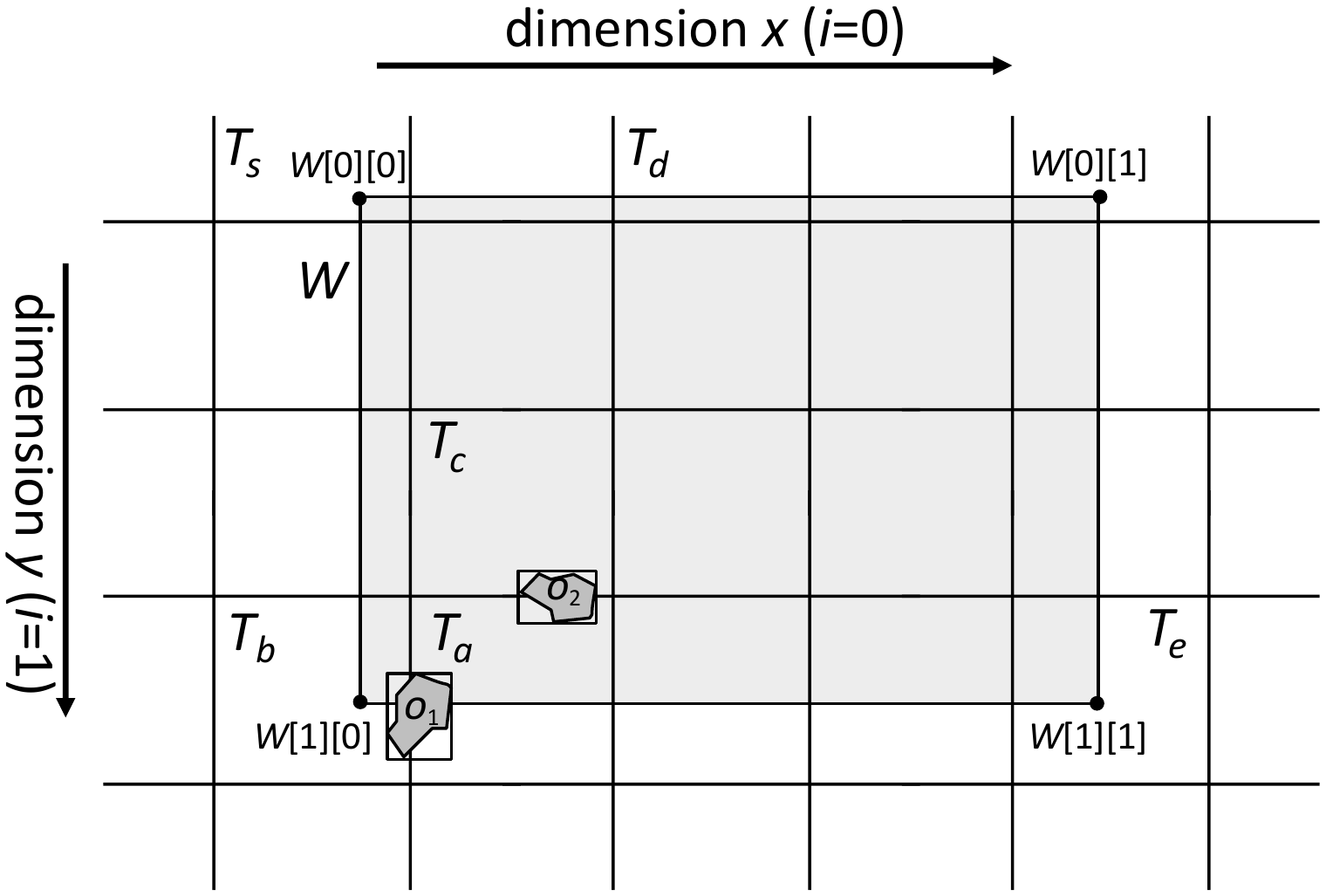}
  \caption{Example of tiling and query evaluation}
  \label{fig:index1}
\end{figure}

Recall that each MBR $r$ can be represented by an interval of
values at each dimension.
Let $r[i]=[r[i][0], r[i][1]]$ be the projection of rectangle $r$ on the $i$-th axis.
%Let $[r.x_l,r.x_u]$ and
%$[r.y_l,r.y_u]$ be the projections of rectangle $r$ on the $x$ and
%$y$ axis, respectively.
%We also use $r[i]$ to denote the projection of rectangle $r$ in the
%$i$-th dimension and
%In addition, let $r[i][0]$, $r[i][1]$ the lower-end and upper-end
%of projection $r[i]$.
For example, in the 2D space, $r[0][1]$ denotes the
upper bound of rectangle $r$ on dimension $0$ (i.e., the $x$-axis).
Similarly, we use $T[i]=[T[i][0], T[i][1]]$ to denote the projection of a tile $T$ to the
$i$-th dimension.
%Let $[T.x_l,T.x_u]$ and $[T.y_l,T.y_u]$ be
%the corresponding projections of a tile, which are also
%denoted by $T[0]$ and $T[1]$, respectively.
%\fix{Consider dropping notation $x_l$ etc, as it is now redundant}
Given a tile $T$ and a dimension $i$, we use $prev(T,i)$ to denote the
tile $T'$ which is right before $T$ in dimension $i$ and has exactly
the same projection as $T$ in the other dimension(s).
For example, in Figure \ref{fig:index1}, $T_b=prev(T_a,0)$.
$prev(T,i)$ is not defined
%in dimension $i$
for tiles $T$ which are in the first column (for $i=0$)
or row (for $i=1$) of the grid.

%\begin{figure}[htb]

Given a range
query window $W$,
a tile that does not intersect $W$ does not contribute any
results to the query. Specifically, the only tiles $T$ that may contain
query results are those for which $T[i][1]\ge W[i][0]$ and $T[i][0]\le
W[i][1]$ at every dimension $i$ and can easily be enumerated after
finding the tiles $T_s$ and $T_e$, which contain
$W[0][0]$ and $W[1][1]$, respectively.%
\footnote{$T_s$ and $T_e$ can be found in $O(1)$ by algebraic
  calculations if the grid is uniform.} 
Figure \ref{fig:index1} illustrates a window query $W$ in lightgrey color
and its four corner points $W[0][0]$, $W[0][1]$, $W[1][0]$, $W[1][1]$.
The tiles which are relevant to $W$ are between (in both dimensions)
the two tiles $T_s$ and $T_e$.%
\footnote{We conventionally assume that the $x=0$ dimension is from
      left to right and the $y=1$ dimension is from top to bottom.}
%that contain
%$W[0][0]$ and $W[1][1]$, respectively.%

For each tile which is totally covered by the query range
in at least one dimension (e.g., $T_a$ in dimension $0$),
we know that the objects in it certainly intersect $W$ in that dimension. For a
tile $T$ that partially overlaps with $W$ in both dimensions (e.g., $T_b$), we need to
iterate through its objects list to verify their intersection with $W$.
We first check whether the MBR of the object intersects $W$ and then
we might have to verify with the exact geometry of the object at a
refinement step.

An important issue is that neighboring tiles may intersect $W$ and
also contain the same object $o$. In this case, $o$ will be reported
more than once, so we need an approach for handling these duplicates.
For example, in
Figure \ref{fig:index1}, object $o_1$ could be
reported both by $T_a$ and by $T_b$.
%\fix{extend example}
A solution to this problem
%(originally proposed for spatial joins \cite{DittrichS00}),
is to report an object $o$ only at the tile which is before all tiles
(in both
dimensions) where $o$ is found to intersect $W$.
For example, in
Figure \ref{fig:index1}, $o_1$ is 
reported by $T_b$ only, which is before $T_a$.
% \fix{extend example}
An easy approach to perform this test is to compute the 
intersection between the query window and the rectangle and report the
result only if a {\em reference point} of the intersection (e.g., the
smallest value in all dimensions)
is included in the tile \cite{DittrichS00}.
While this solution prevents reporting duplicates, 
it requires extra comparisons
and it is unclear how to apply it 
%it may require
%extra comparisons or bookkeeping and potentially synchronization between the thread%s
%that process the query at the different tiles (a tile must know if
%there are other tiles before it wherein the same object is a query result).
%In addition, it is unclear how such an approach can be applied 
for
non-rectangular range queries.
An alternative and more general
approach is to add the results from all tiles in a hash table, which
would prevent the same rectangle from being added multiple times.
%However, this method is much more expensive than the reference point approach. 
% and for other spatial queries besides
%selections and joins \fix{we must be sure that our approach can be
%  applied in all cases}.
%\fix{mention Dittrich test}

%\fix{Show what are the necessary comparisons that should be applied to
%  determine a result in this simple scheme, in order to later compare
%  this case to our proposal.}

\subsection{A Second Level of Partitioning}
\label{sec:index:classes}
We now present our proposal for improving this basic spatial indexing
approach by introducing a second level of partitioning to the contents
of each tile. Our approach avoids the generation of duplicate results
overall and, hence, it does not require any duplicate
elimination. We propose that the set of MBRs in each tile is further divided into
four classes $A$, $B$, $C$, and $D$
(which are physically stored separately in memory).
Specifically, consider a rectangle $r$ which is assigned to (i.e.,
intersects) tile $T$.
\begin{itemize}
\item
  $r$ belongs to class $A$, if for every dimension
    $i$, the begin value $r[i][0]$ of $r$ falls into projection
    $T[i]$, i.e., if $T[i][0]\le r[i][0]$.
%    point ($r[0][0],r[1][0]$ 
%    its top-left corner%  
%      is
 %     contained in $T$; formally, if $T[0][0]\le r[0][0]\ge T.x_l$ and $r.y_l\ge T.y_l$.}
  \item
    $r$ belongs to class $B$ if its $x$-projection $r[0]$
       begins inside $T[0]$, but its $y$-projection $r[1]$ begins before
       $T[1]$, i.e.,  if $T[0][0]\le r[0][0]$ and $T[1][0]> r[1][0]$.
 %      formally, if $r.x_l\ge T.x_l$ and $r.y_l< T.y_l$.}
     \item
       $r$ belongs to class $C$ if its $x$-projection $r[0]$
       begins before $T[0]$, but its $y$-projection $r[1]$ begins inside
       $T[1]$, i.e.,  if $T[0][0]> r[0][0]$ and $T[1][0]\le r[1][0]$.
       %formally, if $r.x_l< T.x_l$ and $r.y_l\ge T.y_l$.}
     \item
       $r$ belongs to class $D$ if both its $x$- and $y$-projections
       begin before $T$, i.e., if $T[0][0]> r[0][0]$ and $T[1][0]> r[1][0]$.
       %; formally, if $r.x_l< T.x_l$ and $r.y_l< T.y_l$.}
\end{itemize}
We can refer to each class by two bits, one for each dimension. The 
bit in each dimension
indicates whether the rectangle starts before the tile in that
dimension.
Hence, class $A$ can also be referred to as class $00$ because a rectangle
in class $A$ of a tile $T$ does not start before $T$ in both
dimensions. Similarly, classes $B$, $C$, and $D$ can be denoted by
$01$, $10$, and $11$, respectively. This notation can generalized to
an arbitrary number of dimensions $m$, where there are $2^m$ classes
of bounding boxes in each multidimensional tile and $m$ bits are used
to denote each class.

\begin{figure}[t]
\centering
  \includegraphics[width=0.99\columnwidth]{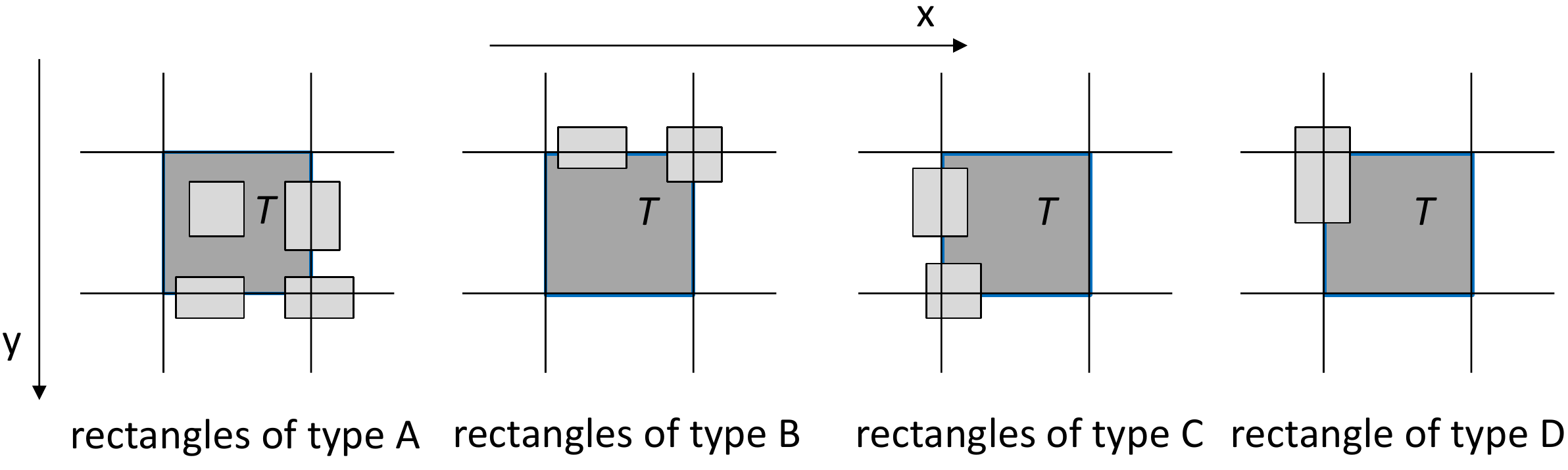}
  \caption{The 4 classes of rectangles inside a tile $T$.}
  \label{fig:classes}
\end{figure}

Figure \ref{fig:classes} illustrates examples of rectangles in a tile
$T$ that belong to the four different classes.
During data partitioning,
for each tile $T$ a rectangle $r$ is assigned to,
we identify its class and,
hence, for each tile, we have four different
rectangle divisions (which are stored separately).
Note that a rectangle can belong to class $A$
of just one tile, while it can belong to other classes (in other
tiles) an arbitrary number of times.

%\begin{figure}[htb]

\section{Range Query Evaluation}
\label{sec:range}
In this section, we show how the divisions can be used to
evaluate spatial range queries efficiently and at the same
avoid generating and
testing duplicate query results.
%We begin our discussion with the most
%common spatial query class: the spatial {\em range} query.
For simplicity, we first consider rectangular range queries where the
query range is a rectangle (window) $W$ and the objective is to find
the objects which spatially intersect $W$. The cases of other query
shapes
% and alternative selection predicates
will be discussed later on.
For now, we focus on the filtering step of the query, i.e., the
objective is to find the object MBRs which intersect $W$.
The refinement step will be discussed in Section \ref{sec:refine}.

%Assume, for now, that $W$ spans more than one tiles per dimension (the
%other cases will be discussed later).

Recall that the tiles which are relevant to $W$ are between the two tiles that contain
$W[0][0]$ and $W[1][1]$ in both dimensions.
We lay out a set of rules that can be used to determine which
rectangles in each of these tiles are query results and what are the
necessary comparisons for determining whether a rectangle is a result.
Finally, these rules can help
us to avoid generating and eliminating duplicate query results,
without any comparisons, bookkeeping, or
synchronization in the processing at different tiles.
%In a nutshell, an object $o$ which intersects $W$
%is reported only at the tile which contains point $(MBR(o)[0][0],
%MBR(o)[1][0])$
%and this does not require any comparisons, bookkeeping, or
%synchronization in the processing at different tiles. 
In summary, the goal of our method is twofold: (i)
eliminate any dependencies between processing at different
    tiles and (ii)  minimize the cost of processing at each tile, by avoiding
     redundant comparisons and duplicate result checks.

\subsection{Selecting relevant classes}
Recall that in order for two rectangles to intersect (in our case $W$
and a candidate query result), they should intersect in {\em all}
dimensions.
In order words, if a rectangle $r$ does not intersect $W$ in one
dimension (i.e., in a dimension $i$, we either have $r[i][1]<W[i][0]$
or $W[i][1]<r[i][0]$), then $r$ is not a query result.
We now present a lemma which can help us to determine
classes of rectangles in a tile that should not be considered in a
query, otherwise they would produce duplicate results.  
%To simplify our
%exposition, our rules determine if a rectangle $r$ in a tile $T$ intersects $W$ in a
%dimension $i$, i.e., if $r[i]$ intersects $W[i]$. By repeating one or more rules
%in each dimension, we can eventually determine whether the rectangles
%$r$ in $T$ are results of a query $W$.
%We present these rules as lemmas that can be trivially
%proved and build up our overall solution.

\begin{lemma}[Filtering]
  \label{lemma:filtering}
  If $W$ intersects tile $T$ and starts before $T$ in dimension $i$, then:
  \begin{itemize}
%    \item all rectangles in $T$ intersect $W$ in dimension $i$;
    \item in the classes having $1$ in dimension $i$, all rectangles 
      that intersect $W$ are guaranteed to 
      intersect $W$ also in the previous tile $prev(T,i)$
      hence they can
      be safely disregarded;
    \item if also $W$ starts before $T$ in dimension
      $j\ne i$, then all objects in the class having $1$ in dimension
      $j$ are guaranteed to 
      intersect $W$ also in the previous tile $prev(T,j)$
      hence they can
      be safely disregarded.
  \end{itemize}
\end{lemma}

To understand the first point of the lemma, consider
again Figure \ref{fig:index1} and tile $T_a$;
$W$ starts before $T_a$ in dimension $0$.
All rectangles of $T_a$ in classes $C=10$ and
$D=11$ can be ignored by tile $T_a$ when processing query $W$ because
these rectangles are guaranteed to also intersect the previous tile
$T_b=prev(T_a,0)$ in dimension $0$ and they can be processed there.%
\footnote{Note that if $W$ also covers $T_b$ at dimension
$0$, a rectangle $r$ in $T_b$ will be processed 
recursively at $prev(T_b,0)$, if $r$ is in a class of $T_b$ having $1$ in dimension $i$.}
Hence, $o_1$ (which belongs to class $C=10$) is not examined at all by
tile $T_a$.

To understand the second point of the lemma, consider
object $o_2$ in Figure \ref{fig:index1}.
Note that $W$ also starts before $T_a$ in dimension $1$. 
%Tile $T_a$ is totally covered
%by $W$ in dimension 0 and $W$ begins before $T_a$ in dimension $1$.
This guarantees that all objects in class $B=01$ of tile  $T_a$
which intersect $W$ can be reported at tile
$T_c=prev(T_a,1)$, so $T_a$ can safely ignore all rectangles in class
$B=01$.

\subsection{Reduction of comparisons}     

We now turn our attention to minimizing the comparisons needed for
rectangle classes that {\em  have to} be checked
(i.e., those not eliminated by Lemma
\ref{lemma:filtering}).
If a tile $T$ is covered by the window $W$ in a dimension $i$,
then we do not have to perform intersection tests in that dimension.
Let us go back to the example of Figure \ref{fig:index1}.
Recall that only rectangles in class $A=00$ of tile $T_a$
need to be checked against window $W$, because the other classes have
been filtered out by Lemma \ref{lemma:filtering}.
For all these rectangles,
we only have to
conduct an intersection test with $W$ in dimension $1$,
since $T_a$ is totally covered by $W$ in dimension $0$.
For the dimension(s) where the tile is not covered by $W$,
the following lemmas can be used to further reduce the necessary comparisons.
%The first point of Lemma \ref{lemma:filtering} indicates that
%for any rectangle $r$ in $T$ which needs to be compared to $W$, it
%suffices to perform the comparison in dimension $j\ne i$.
%Obviously, if $W$ covers $T$ also in dimension $j$ (e.g., tile $T_c$
%in Figure \ref{fig:index1}), there is no need for any comparison.

%The following lemma applies to classes of rectangles in each tile $T$
%that {\em have} to be checked (i.e., those not eliminated by Lemma
%\ref{lemma:filtering}).

\begin{lemma}[Comparisons Reduction 1]
  \label{lemma:comp1}
  If $W$ ends in tile $T$ and starts before $T$ in dimension $i$, then
  for a rectangle $r\in T$, $r$ intersects $W$ in dimension $i$ iff $r[i][0]\le W[i][1]$. 
\end{lemma}
For example, in tile $T_a$, we only have to test intersection in
dimension $1$ for rectangles $r$ in class
$A=00$, as already explained. The intersection test can be
reduced to a simple comparison, i.e., if $r[1][0]\le W[1][1]$ then $r$
intersects $W$.
Symmetrically, we can show:
\begin{lemma}[Comparisons Reduction 2]
  \label{lemma:comp2}
  If $W$ starts in tile $T$ and ends after $T$ in dimension $i$, then
  for a rectangle $r\in T$, $r$ intersects $W$ in dimension $i$ iff $r[i][1]\ge W[i][0]$. 
\end{lemma}
For example, consider tile $T_d$ in Figure \ref{fig:index1}.
Due to Lemma \ref{lemma:filtering}, we can eliminate from
consideration rectangle classes $C=10$ and $D=11$ in $T_d$, while for
the rectangles in classes $A=00$ and $B=01$, the rectangles are
guaranteed to intersect $W$ in dimension $0$. Hence, we only have to
find the rectangles $r$ in classes $A$ and $B$, for which
$r[1][1]\ge W[1][0]$.

\eat{
process all rectangles $r$ in class
$A=00$ to confirm if $r[1][0]\le W[1][1]$. If these rectangles are
ordered by $r[1][0]$, the number of comparisons can further be
reduced.

Lemma \ref{lemma:comp1} suggests that if the rectangles are ordered by
$r[i][1]$, we can reduce the necessary comparisons in a class where
the intersection test is necessary only in dimension $i$, by
conducting a search (e.g., binary search) for the first object that
satisfies  
$r[i][1]\ge W[i][0]$.

Lemma \ref{lemma:comp2} is symmetric to Lemma  \ref{lemma:comp1}.
For example, in tile $T_a$, we only have to process all rectangles $r$ in class
$A=00$ to confirm if $r[1][0]\le W[1][1]$. If these rectangles are
ordered by $r[1][0]$, the number of comparisons can further be
reduced.
}

\begin{figure}[t]
\centering
  \includegraphics[width=0.75\columnwidth]{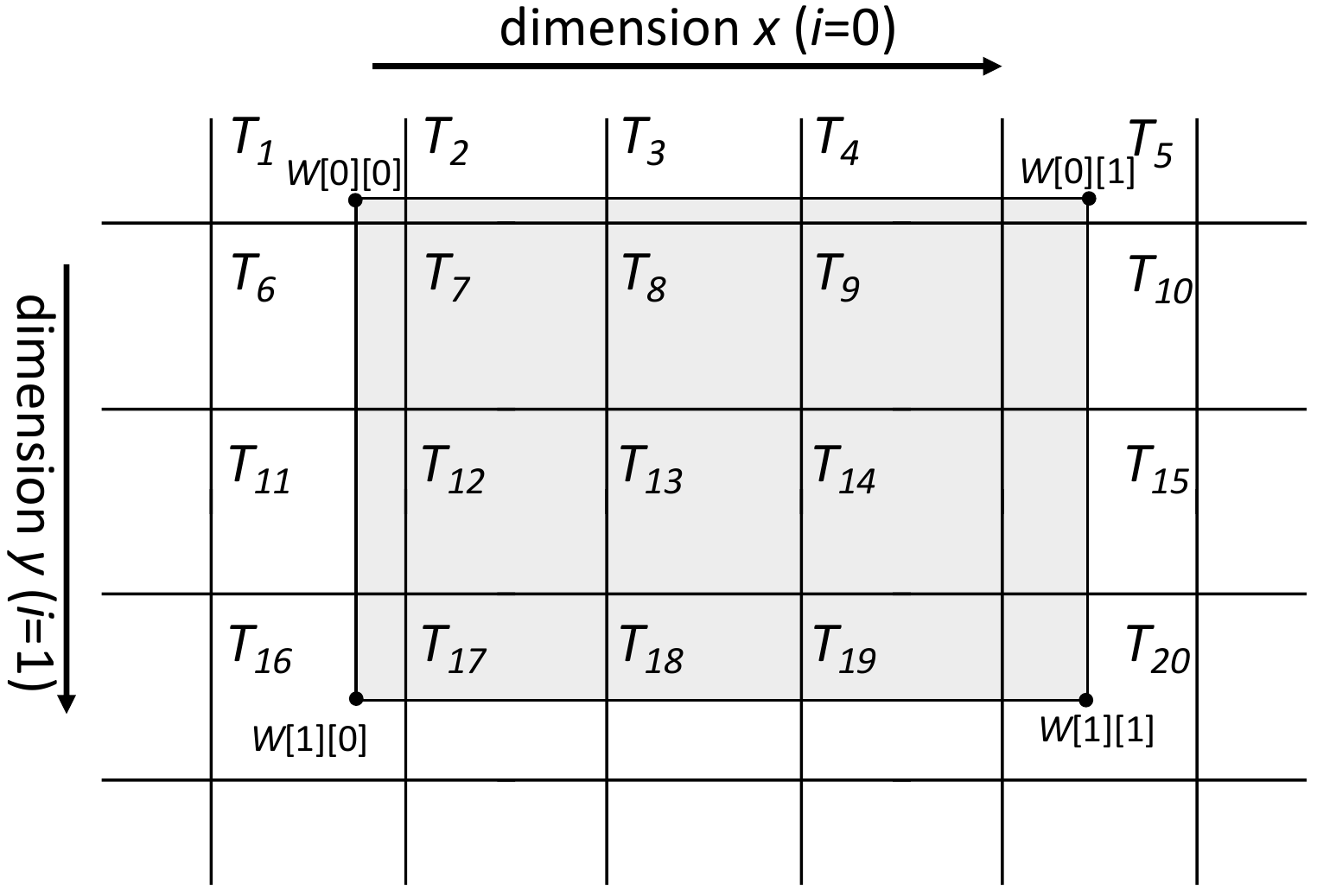}
  \caption{Processing a range query at each tile}
  \label{fig:tile_processing}
\end{figure}

\stitle{Example.} A detailed example of the tasks executed by each tile in a window query
$W$ is illustrated in Figure \ref{fig:tile_processing}.
Tile $T_1$ processes all four classes of its rectangles.
For each rectangle, just one comparison is needed per dimension due to
Lemma \ref{lemma:comp2}.
%If the rectangles are sorted by their end
%point in a dimension, this can be used to accelerate finding the
%qualifying rectangles.
Tiles
$T_2$--$T_5$ process only classes $A=00$ and $B=01$ due to Lemma
\ref{lemma:filtering}.
In $T_2$--$T_4$, the intersection test at dimension $0$ is skipped. In
addition,
for each rectangle, only one comparison is necessary (Lemma
\ref{lemma:comp2}).
%if the data in each class are ordered by their end
%point in dimension $1$, this can be exploited (Lemma
%\ref{lemma:comp1}).
Tile $T_5$ applies one comparison in each dimension using the
start and end points in dimensions $0$ and $1$, respectively
(Lemmas \ref{lemma:comp1} and \ref{lemma:comp2}).
For tile $T_6$ only rectangle classes $A=00$ and $C=10$ need to be
processed and Lemma \ref{lemma:comp2} can be used to reduce the
comparisons for dimension $0$, while there is no need for comparisons
in dimension $1$, as $W$ covers the tile in this dimension.
For tiles $T_7$--$T_{10}$, only rectangle classes $A=00$ are accessed.
For tiles $T_7$--$T_{9}$, no comparisons at all are required, whereas
for tile $T_{10}$, one comparison for dimension $0$ should be
performed (Lemma \ref{lemma:comp1}).
Tiles $T_{11}$--$T_{15}$ are processed as tiles $T_6$--$T_{10}$
(respectively).
Tile $T_{16}$ processes classes $A=00$ and $C=10$ only and two
comparisons for each rectangle are required.
Tiles $T_{17}$--$T_{20}$ process only class $A=00$.
In Tiles $T_{17}$--$T_{19}$, one comparison per rectangle is required
(Lemma \ref{lemma:comp1}), while Tile $T_{20}$ requires two
comparisons per rectangle.

%\begin{figure}[htb]

\subsection{Overall approach}

Given a range query $W$,
we first identify the range of tiles that intersect $W$ (i.e., the
first and the last tile in each dimension) by simple algebraic
operations (i.e., by dividing the endpoints of $W$ in each dimension
by the number of space divisions in that dimension).
We then pass the control to each tile $T$, which
accesses the relevant classes of rectangles and perform the necessary
computations for the rectangles in them.
For each qualifying rectangle a refinement step is performed (after accessing the
corresponding object's geometry).
The results produced at each tile
are eventually merged.

\subsection{Disk queries}

We now discuss the evaluation of disk range queries,
%queries, which do not have
%rectangular shape. We begin by the {\em disk} query,
where the objective
is to find all objects which overlap a disk $D$ of radius $\epsilon$ centered at
a given point $q$. This query is equivalent to a {\em distance range} query of
the following form: ``find all objects having distance at most $\epsilon$
from location $q$'' and it is very popular in location based services applications. 

To evaluate a disk query on our two-level partitioned dataset, we
apply a similar method, as for the rectangular window queries that we have seen
already; we first find the tiles that intersect with the disk and then
the objects in them that satisfy the query predicate. If we
approximate the disk $D$ by
%its MBR
$MBR(D)$, we can easily identify the tiles that
potentially intersect the disk by simple algebraic computations, as in
window queries.
For each such tile, its {\em minimum distance} to $q$ is computed and, if
the distance is found at most $\epsilon$, the tile is confirmed to intersect the
disk.%
\footnote{In practice, we do not have to compute the minimum distance
  for each tile that intersects  $MBR(D)$.
  For each row of tiles intersecting  $MBR(D)$, we can just find the first tile $T_s$
  with distance at most $\epsilon$ to $q$, by scanning the row forward,
  and then the  last tile $T_e$  with distance at most $\epsilon$, by
  scanning the row backward. All tiles between $T_s$ and $T_e$ are
  guaranteed to qualify the minimum distance predicate.}

We turn our attention to computing results in a tile and to duplicate
avoidance.
Since a rectangle
can be assigned to multiple tiles, the objective is to examine only
the MBR classes in each tile which are necessary to ensure that (i) no
result is missed and (ii) no duplicate results are reported. In other
words, all rectangles that intersect the disk should be reported
exactly once. For this, we follow a similar approach as for the case of
rectangular queries. For each tile $T$, where $T$ is within distance
$\epsilon$ from $q$, we check whether $prev(T,i)$ in each dimension $i$ is
also within distance $\epsilon$ from $q$, i.e., whether $prev(T,i)$
is in the set of tiles $S$ that may include results. If yes, then we
disregard the corresponding class of rectangles in $T$. Hence, if
$prev(T,0)\in S$, then class $B=01$ is disregarded, whereas if
$prev(T,1)\in S$, then class $C=10$ is disregarded. If
$prev(T,0)\in S$ and $prev(T,1)\in S$, then all classes $B,C,D$ are
disregarded.

\begin{figure}[t]
\centering
  \includegraphics[width=0.9\columnwidth]{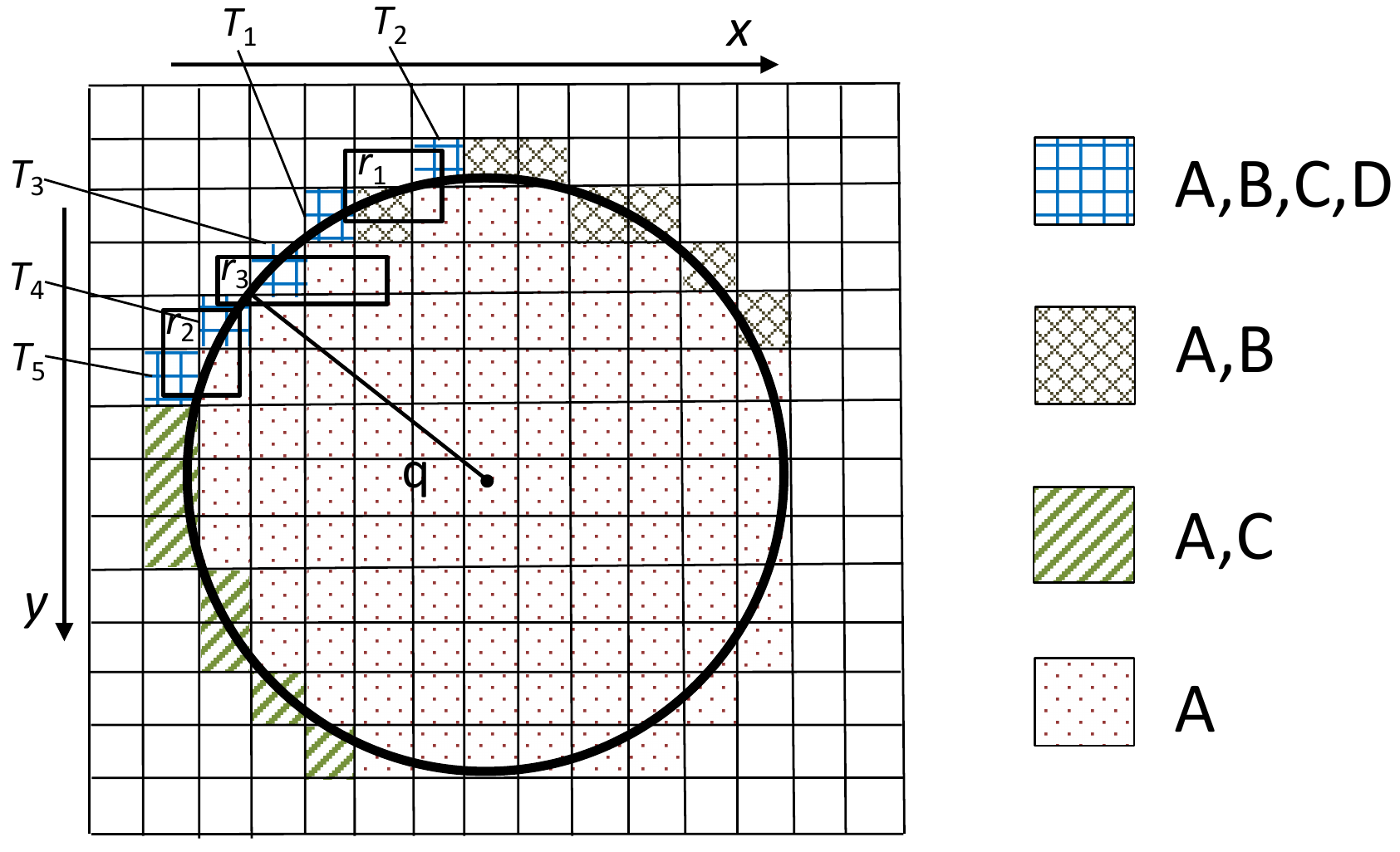}
  \caption{Example of disk query evaluation}
  \label{fig:disk}
\end{figure}

Figure \ref{fig:disk} shows an example of a disk query centered at
q. The tiles which intersect the disk are shown by different patterns
depending on the classes of rectangles in them that have to be
checked. For example, in tile $T_5$ all four classes will be examined
(we call $T_5$ an $ABCD$ tile, in the context of the disk query).
Note that for the majority of tiles which intersect the disk range, we
only have to examine rectangles in class $A=00$.

A subtle point here is that if we simply examine all
rectangles in the classes that correspond to each tile, we may end up
examining duplicates. For example, consider rectangle $r_1$, which
will be examined in both tiles $T_1$ (in class $B$) and $T_2$ (in class
$C$). To avoid such duplicates, for each rectangle in an $ABCD$ tile $T$, 
if the tile is closer to $q$ in the $y$-dimension compared to the
$x$-dimension, we ignore rectangles $r$ in classes $C$ and $D$,
for which $r[1][1]>T[1][1]$ (these will be handled in another tile).
For example, in tile $T_2$, we ignore rectangles in classes $C$ or
$D$, which ``overflow'' to the tile below $T_2$ (such as $r_1$). 
If the tile is closer to $q$ in the $x$-dimension compared to the
$y$-dimension, we ignore rectangles $r$ in classes $B$ and $D$,
for which $r[0][1]>T[0][1]$.
For example, in tile $T_5$, we ignore rectangles in classes $B$ or
$D$, which ``overflow'' to the tile on the right of $T_5$ (such as $r_2$). 
Finally, for a single tile, which has (almost) the same distance to
$q$ in both dimensions, we consider all rectangles, regardless whether
they ``overflow'' or not to the next tiles. For example, for tile
$T_3$ we consider all rectangle classes, i.e., rectangle $r_3$ will be
examined in $T_3$ and not in $T_4$.

Before examining the rectangles in the tiles in $S$ (only the relevant
classes), we can compute the {\em maximum distance} between the tile and
$q$, and if this distance is found to be at most $\epsilon$, then the tile is
marked as {\em covered by} the disk. For tiles which are covered by
the disk, we do not have to verify any distances between the objects
assigned to them and $q$, as these distances are guaranteed to have
distance at
most $\epsilon$ from $q$ (i.e., they are definite results). Again, at each row, the
set of tiles which are covered by the disk are continuous, meaning
that we only have to check the tiles in both directions starting from
the tile which includes $q$ in dimension $x$ until we find the first
one that violates the maximum distance condition.   

Finally, the method described above for disk queries can be
generalized for any query whose range is a convex polygon.
%(i.e., all range queries commonly applied).
We first find the set of tiles $S$ which
intersect the query range. Then, for each tile $T\in S$, we determine which
classes of objects need to be examined (i.e., exclude classes that
would produce duplicates). For each tile which is totally
covered by the query region, we just report its contents in the
relevant classes as results and for the remaining tiles we conduct an
intersection test for each rectangle before determining whether it is a result.

\section{Refinement Avoidance}
\label{sec:refine}
We now discuss the evaluation of the refinement step of range queries
on our two-level partitioning index.
%\fix{give a name to our index? e.g. 2LP}
We begin by a general and important lemma,
which applies independently to our index and
greatly reduces the number of objects for which the refinement
step needs to be applied, especially for query ranges which are
relatively large.

\begin{lemma}[Refinement Step Avoidance]
  \label{lemma:ref}
  Given a candidate object whose MBR $r$ intersects the query range,
  if at least one side of $r$ is inside the query range, then the
  object is guaranteed to intersect the range and no refinement step
  is necessary.
\end{lemma}

The lemma is trivial to prove, based on the definition of MBR. Recall
that the MBR of an object is defined by the minimum and maximum values
of the object in every dimension. Hence, at each side of the MBR,
there is at least one point
which is part of the object's geometry. If one side of the MBR is inside the
query range, then there should be at least one point of the object
inside the query range, i.e., the object and the range intersect.
The lemma generalizes to more than two dimensions. In a
$d$-dimensional space, we test if at least one of the
$(d-1)$-dimensional faces of the minimum hypercube that bounds the
object is inside the query range. 

\begin{figure}[htb]
  \centering
  \subfigure[window query]{
    \label{fig:rskip}
     \includegraphics[width=0.3\columnwidth]{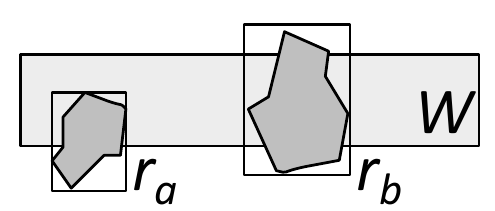}
    }
   \hspace{0in}
  \subfigure[disk query]{
    \label{fig:rskip2}
    \includegraphics[width=0.27\columnwidth]{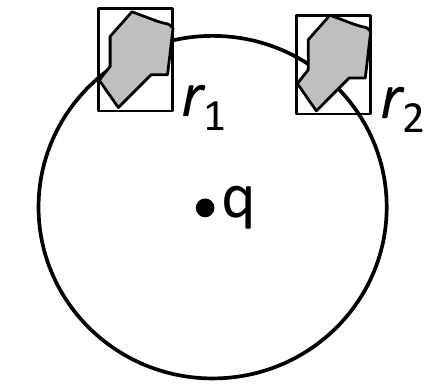}
  }
  \vspace{-0.1in}
  \caption{Refinement step avoidance}
  \label{fig:refinement_avoidance}
\end{figure}

For different range shapes, we can define specialized MBR
side coverage tests.
Consider a rectangular query range $W$ and an object MBR $r$ that
intersects $W$. To apply a refinement avoidance test, we should verify if $W$ covers $r$
in at least one dimension. If this is true, given that $r$ intersects
$W$, one of the cases shown in Figure \ref{fig:rskip}
should happen. Either one side of $r$ is inside the window $W$ and
Lemma \ref{lemma:ref} applies (see $r_a$ in Figure \ref{fig:rskip})
or $r$ splits $W$ along the coverage axis (see $r_b$ in Figure \ref{fig:rskip}).
%In this case an edge of the rectangle is not contained in $W$, but $W$
%splits the MBR along the coverage axis; it is not possible that the
%object in the MBR does not intersect $W$, unless the area that forms
%the object is not connected.
In both cases,
whatever the geometry of the object is the object definitely
intersects $W$.%
\footnote{We assume that the geometry of each object is  continuous.}  
For a disk query range $D$, we can check whether there are at least
two corners of $r$ whose distances to the disk center are smaller than
or equal to the disk radius. For example, in Figure \ref{fig:rskip2},
rectangle $r_1$ has at least two corners in the disk range, which
means that at least one side of the rectangle is in the range and
Lemma \ref{lemma:ref} applies. On the other hand, only one corner of
$r_2$ is inside the disk, hence the refinement step for the
corresponding object cannot be avoided.

Let us now turn our focus to our index and see how we can take
advantage of Lemma \ref{lemma:ref} to apply the refinement avoidance
test and minimize the necessary comparisons.
The main idea is to study the refinement avoidance test at the
{\em tile level}, in order to limit the comparisons required for each 
class of objects in the tiles.

Specifically, for each $T$ that intersects a query range $W$
and for each dimension $i$, we consider two cases:
(i) $W$ starts before $T$ in dimension $i$, i.e.,  $W[i][0]<T[i][0]$
and (ii) $W[i][0]\ge T[i][0]$.
In the first case, due to Lemma \ref{lemma:filtering}, only classes of
rectangles that start inside $T$ in dimension $i$ are considered,
which means for each rectangle $r\in T$ which is found to intersect
$W$, we already know that $W[i][0] <r[i][0]$. Hence, we only have to
test if $r[i][1]\le W[i][1]$ to confirm that $r$ is covered by $W$ in
dimension $i$ and that the refinement step for $r$ is not necessary.
On the contrary, for the case where $W[i][0]\ge T[i][0]$, we should
apply the complete refinement avoidance test
in dimension $i$ (i.e.,  $W[i][0] \le r[i][0]$
and $r[i][1]\le W[i][1]$) for each $r\in T$ which is found to intersect
$W$.

As an example, consider again the query in Figure
\ref{fig:tile_processing}.
For each rectangle in tile $T_1$ found to intersect $W$, we should
apply the complete coverage test in each of the two dimensions, before
applying the refinement step.
For each rectangle $r$ in tiles $T_2$--$T_5$ and for dimension $0$ we
only have to test if  $r[0][1]\le W[0][1]$, since these rectangles are
in classes $A$ and $B$ and they start inside the tile in dimension $0$.
Similarly, for each rectangle $r$ in tiles $T_6$, $T_{11}$, and
$T_{16}$ and for dimension $1$ we only have to test
if  $r[1][1]\le W[1][1]$, since these rectangles are
in classes $A$ and $C$ and they start inside the tile in dimension $1$.
For all remaining tiles, if $r[0][1]\le W[0][1]$ or
$r[1][1]\le W[1][1]$ holds, $r$ is guaranteed to be a true result and
no refinement is necessary.

\eat{
Recall that the only objects for which a refinement step is necessary
are those for which $W$ does not cover their MBR in at least one dimension.
The only possibility for this to happen is that the MBR contains a
corner of $W$.  The refinement step can be simplified by executing two
{\em ray tracing} operations \fix{citation needed}
from the corner of $W$ inside the MBR to the
directions of the edges that form the corner (see Figure
\ref{fig:rexec}). If at least one of these
operations hits the object the object is a result.
\fix{mention complexity}

\fix{investigate additional rules for convex objects(?)}
\fix{investigate additional approximations for objects in classes(?)}
}

\section{Batch Query Processing}
\label{sec:parallel}
In the previous sections, we presented how our two-level index handles single query requests.
Real systems however receive and need to evaluate a large number of concurrent queries.
Under this, we next discuss how to efficiently process batches of spatial range queries.
Although our focus is primarily in a single-threaded processing environment,
parallel query processing in modern multi-core hardware can also benefit from
the ideas discussed in this section. To this end, our experimental analysis includes
both single-threaded and multi-threaded experiments.

Naturally, a straightforward approach for processing a workload of concurrent spatial
range queries is to evaluate every query independently, directly applying the ideas
discussed in the previous sections. In a parallel processing environment, we can
easily adopt this approach by assigning the queries to the available threads
in a round robin fashion.
We call this simple approach \qatomic.
Its main shortcoming is that it is cache agnostic; 
%Despite its simplicity this approach exhibits a strong shortcoming.
as every issued query $q$ typically overlaps multiple tiles of the grid,
the computation of $q$ requires accessing data structures
in different parts of the main memory,
i.e., the memory access pattern is prone to cache misses.
% This cache agnostic approach of \textsf{queries-atomic} computation creates a main memory access pattern prone to cache misses.
The problem is present also in parallel query processing,
as every thread goes through multiple rounds of ``content switching''.

To address this shortcoming of \qatomic, we design a
cache-conscious two-step approach.
%based on the following two-step approachprinciple.
Given a large batch of queries $Q$, for each tile, accumulate the subtasks
of all queries in $Q$ that intersect the tile.
Each subtask corresponds to accessing and processing (the relevant to the query)
classes of rectangles in the tile.
Then, in a second step, we initiate one process at each tile, which evaluates
the corresponding subtasks.
%Instead of scheduling 
%When the data structures for a tile are accessed in main memory,
%we evaluate in batch the subtasks of every query related to $T$.
Essentially, query processing is no longer driven by the queries, but from the grid tiles and therefore, we call this approach \tatomic.
%Given a batch $Q$, the evaluation of the queries involves two steps. First, we identify the range of tiles that intersect the area (window or disk) of every query $q \in Q$. This information is stored 
This method is favored by parallel processing,
since each thread (corresponding to a tile)
can benefit from the processor's cache
while processing the subtasks assigned to it.
As we demonstrate in Section \ref{sec:exps} the 
\tatomic
approach scales better with the number of parallel threads
compared to \qatomic.

\section{Experimental Evaluation}
\label{sec:exps}
In this section we present our experimental analysis.
We first describe our setup and then our experiments,
which investigate the construction and update costs of our
two-level index as well as its efficiency and scalability
in spatial range query evaluation.

\subsection{Setup}
Our analysis was conducted on a machine with
384 GBs of RAM and a dual Intel(R) Xeon(R) CPU E5-2630
v4 clocked at 2.20GHz running CentOS Linux 7.6.1810.
All methods were implemented in C++, compiled using \textsf{gcc} (v4.8.5) with flags \texttt{-O3}, \texttt{-mavx} and \texttt{-march=native}.
For our parallel processing tests, we used OpenMP and activated hyper-threading, allowing us to run up to 40 threads. 

%\eat{
\begin{table}[t]
\centering
\small
\caption{Datasets used in the experiments}
\label{tab:datasets}
\begin{tabular}{|@{~}c@{~}|@{~}c@{~}|@{~}c@{~}|@{~}c@{~}|@{~}c@{~}|}\hline
\textbf{dataset}		&\textbf{type}	&\textbf{card.}	&\textbf{avg. $x$-extent}	&\textbf{avg. $y$-extent}\\\hline\hline
AREAWATER			&polygons		&$2.3$M					&$0.00000723$ 			&$0.00002296$ \\
LINEARWATER 		&linestrings       &$5.8$M					&$0.00002224$ 			&$0.00007319$ \\
ROADS					&linestrings		&$20$M					&$0.00001254$ 				&$0.00004067$ \\
EDGES					&polygons		&$70$M 					&$0.00000610$ 			&$0.00001982$ \\\hline
\end{tabular}
\end{table}

\stitle{Datasets.} We experimented with four of the publicly available Tiger 2015 datasets
\cite{EldawyM15}.%
\footnote{http://spatialhadoop.cs.umn.edu/datasets.html}
The input objects were normalized so that the coordinates in each dimension take values inside $[0, 1]$.
Table~\ref{tab:datasets} provides statistics about the datasets we used.
The datasets contain from 2.3M to 70M objects, either polygons or linestrings. The last two columns of the table are the relative (over the entire space) average length for every object's MBR at each axis.

\stitle{Methods}. We designed two variants of our two-level indexing
approach (presented in Section \ref{sec:index:classes}). In the first
one termed \twolevel, each tile of the  grid stores the (MBR, id)
pairs of the indexed objects in
four tables (one for each of the $A$, $B$, $C$, $D$ classes),
such that there is no particular order of the contents of each table
(i.e., as in a heap file).
This organization supports insertions efficiently as the MBRs of new
objects are simply appended to the tables of the tiles.
In the second variant, termed \twolevelplus, the MBRs of each class are
also stored in four decomposed tables,
following
the Decomposition Storage Model (DSM) \cite{CopelandK85}, adopted by column-oriented database systems (e.g., \cite{StonebrakerABCCFLLMOORTZ05}).
Specifically, each rectangle $r$ with id $i$ is decomposed to four
tuples, i.e., $\langle r[0][0], i\rangle$, $\langle r[0][1],
i\rangle$, $\langle r[1][0], i\rangle$, $\langle r[1][1], i\rangle$
and each tuple is stored in a dedicated table. 
The tables are sorted by their first column and used to evaluate fast
queries on tiles, where just one endpoint of each MBR needs to be compared
(see Lemmas \ref{lemma:comp1} and
\ref{lemma:comp2}).
In this case, \twolevelplus takes advantage of the sorted decomposed
tables to reduce the information that has to be accessed and the
number of comparisons.
For example, in tile $T_2$ of Figure \ref{fig:tile_processing}, we
only have to access the decomposed tables of classes $A$ and $B$
with $\langle r[1][1], i\rangle$ tuples to test whether $r[1][1]\ge
W[1][0]$ for each rectangle there. Since these tuples are sorted, we
perform binary search to find the first qualifying tuple in
each table and then scan the tables forward from thereon.
%and sorted in different ways (by lower and by upper end at each dimension following the idea of the
\twolevelplus processes window queries very fast, but it is suited mostly for static data.

We also considered three competitors to our indexing scheme. The \onelevel scheme, discussed in Section~\ref{sec:index:grid}, indexes the MBRs of the input objects using a simple uniform grid; MBRs that span multiple tiles are replicated accordingly, but
our proposal for a second level of indexing (i.e., Section~\ref{sec:index:classes}) is not applied.
When processing window queries, \onelevel performs duplicate elimination using the reference point approach \cite{DittrichS00}. %Specifically, a rectangle in a tile $T$, which is found to intersect the query window is reported at $T$
%only if the $x$-low, $y$-low point of the intersection area between the rectangle and the window lies inside $T$.  
%its intersection with the query window is inside the same tile as the rectangle.
The second competitor is an in-memory STR-bulkloaded R-Tree \cite{LeuteneggerEL97} taken from the Boost library (boost.org),
which has a fanout of 16 for both inner and leaf nodes.
%and uses the packing algorithm \cite{LeuteneggerEL97,GarciaLL98}.
This configuration is reported to perform the best (we also
confirmed this by testing).
The third competitor is BLOCK; the implementation was kindly provided by the authors of \cite{OlmaTHA17}.
After testing this approach we found it to be orders of magnitude slower that the R-tree
(BLOCK takes seconds to evaluate range queries on our data), which can be attributed to the fact that
BLOCK is implemented for 3D objects.
%, its performance was significantly worse than the rest of the tested methods.
Therefore, we decided to exclude BLOCK from the reported measurements.
\stitle{Tests and parameters}. 
To assess the effectiveness of the tested indices, we compared their space requirements,
their building and update costs and their query performance.
For the partitioning-based schemes, i.e., \onelevel, \twolevel and
\twolevelplus, we investigated the best granularity for their grid by
varying the number of partitions (i.e., divisions) per dimension.
By default, we use 2000 per dimension, resulting in a 2000$\times$2000
grid.
We experimented with
both window and disk queries by varying their relative extent over the
entire data space. We considered both cases of single and batch query
processing, measuring the average execution time per query and the
total execution time, respectively. 
%Especially, for batch query processing
%we also experimented with the number of queries and for
%parallel execution, with the number of available threads.
Especially, for parallel batch query processing, we also experimented with the number of available threads.
At each experimental instance, we averaged the cost of 10000 queries
applied on non-empty areas of the map (i.e., the queries always return
results). By default the areas of the queries are $0.1$\% of the area of the map.
%Table~\ref{tab:parameters} summarizes our experimental parameters and their default values.
%\fix{+++ add the table}
\eat{For range queries, our performance measure is query throughput. Given
a workload of queries report the throughput (queries per second) for
each tested method. 

Parameters: (a) for different query areas plot throughput, (b) for
queries that have same area (e.g. in $km^2$, split them into groups based on
selectivity and report avg. throughput for each group.
 }

\subsection{Filtering Vs Refinement}
\begin{figure*}[t]
\begin{center}
\begin{small}
\fbox{
{\small filtering}
%\hspace{0.5ex}
\includegraphics[width=0.06\columnwidth]{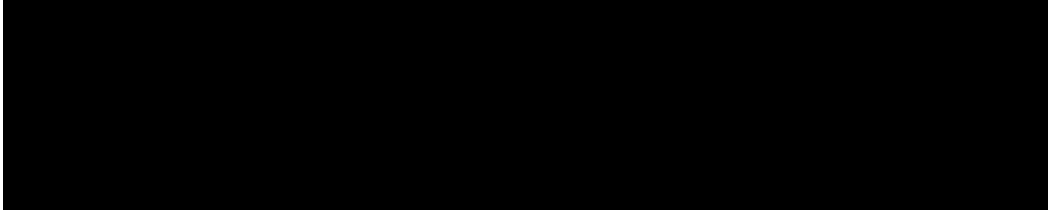}
\hspace{1ex}
{\small avoidance}
%\hspace{0.5ex}
\includegraphics[width=0.06\columnwidth]{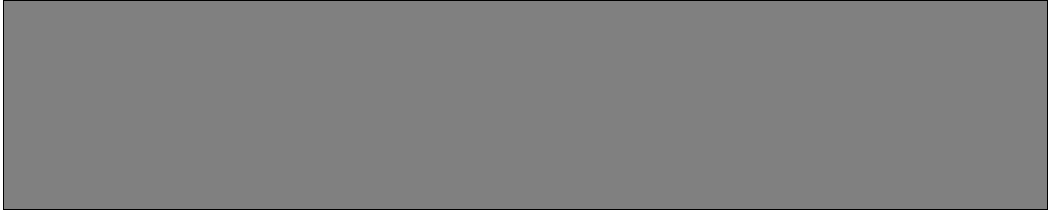}
\hspace{1ex}
{\small refinement}
%\hspace{0.5ex}
\includegraphics[width=0.06\columnwidth]{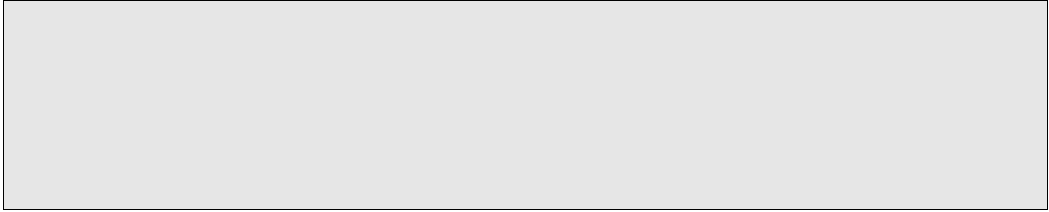}
\hspace{1ex}
}
\end{small}
\end{center}
%\vspace*{-4ex}
\begin{center}
\begin{tabular}{cccc}
AREAWATER &LINEARWATER &ROADS &EDGES\\
\includegraphics[width=0.5\columnwidth]{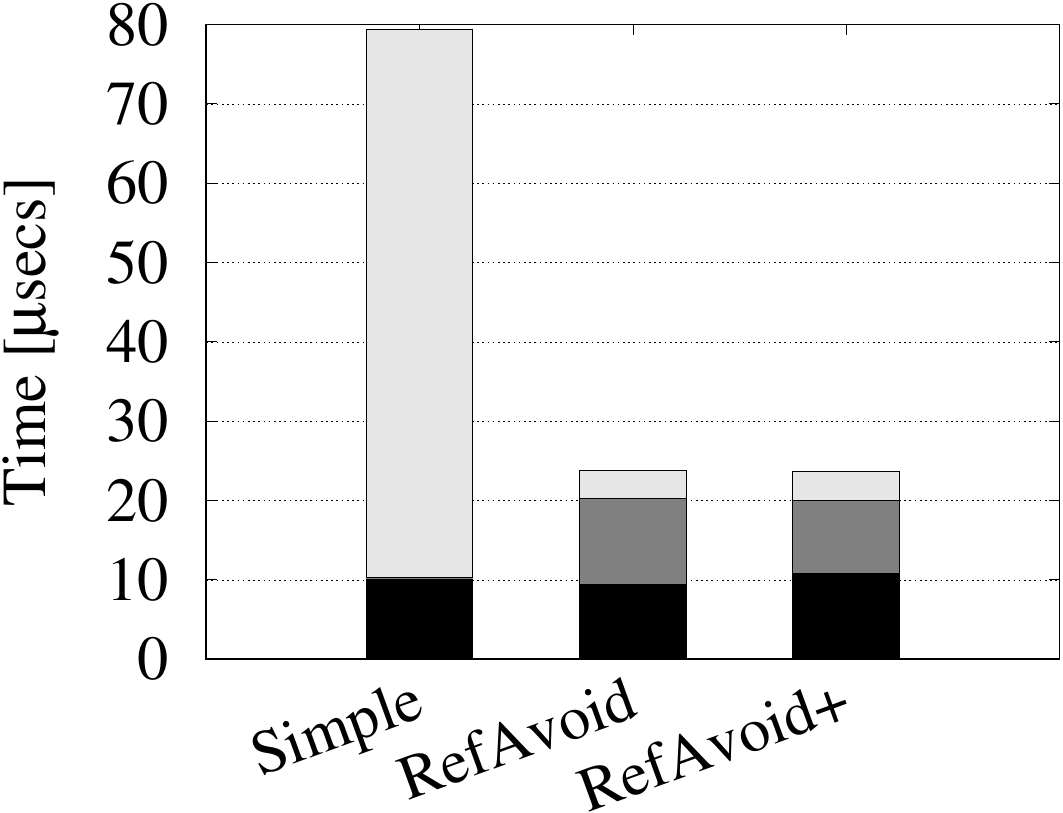}
&\hspace{-1ex}\includegraphics[width=0.5\columnwidth]{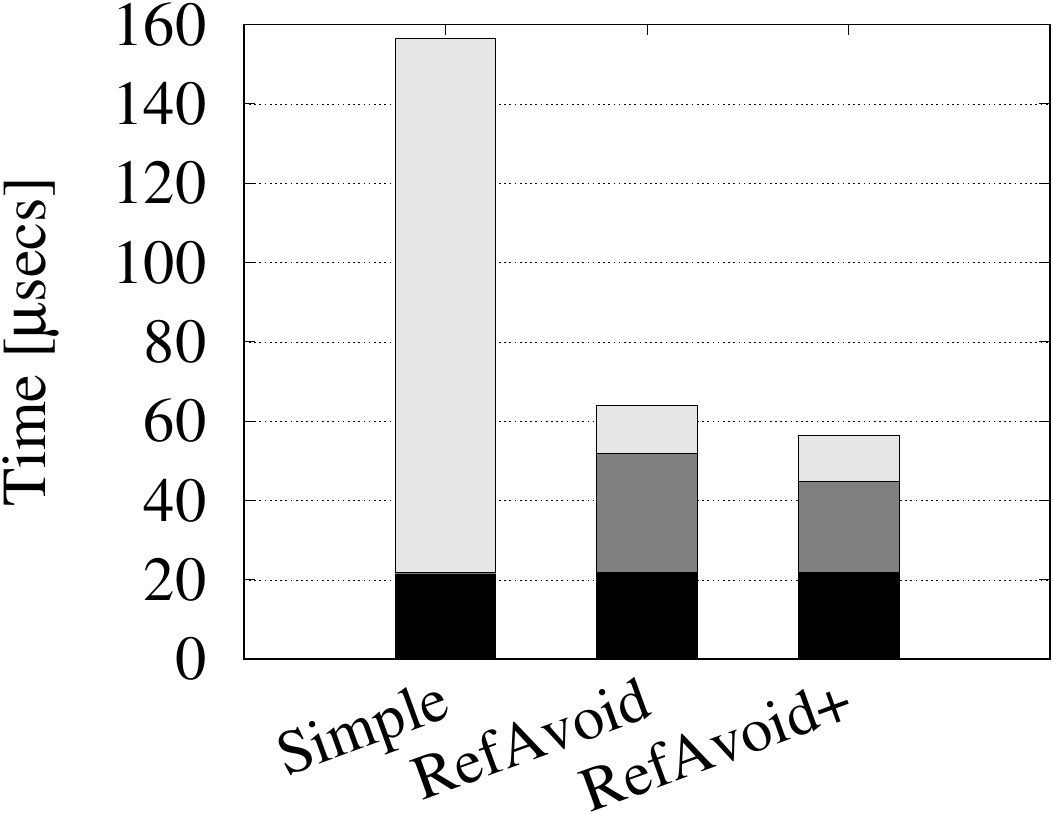}
&\hspace{-1ex}\includegraphics[width=0.5\columnwidth]{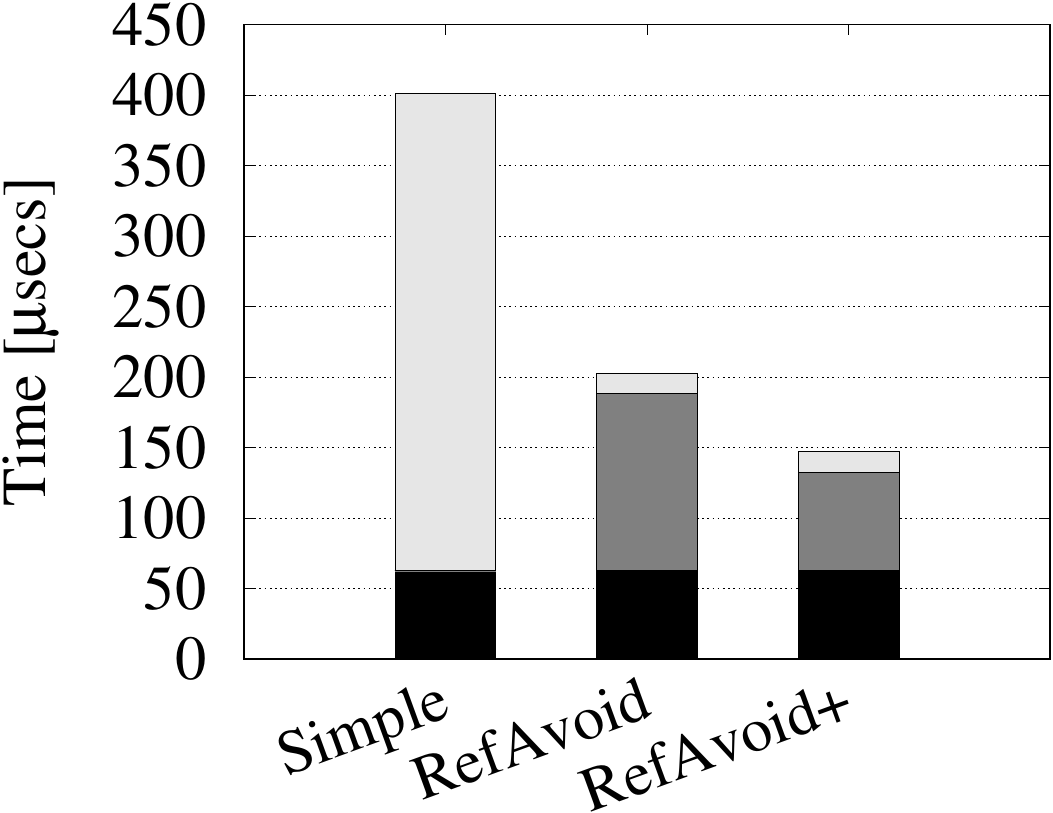}
&\hspace{-1ex}\includegraphics[width=0.5\columnwidth]{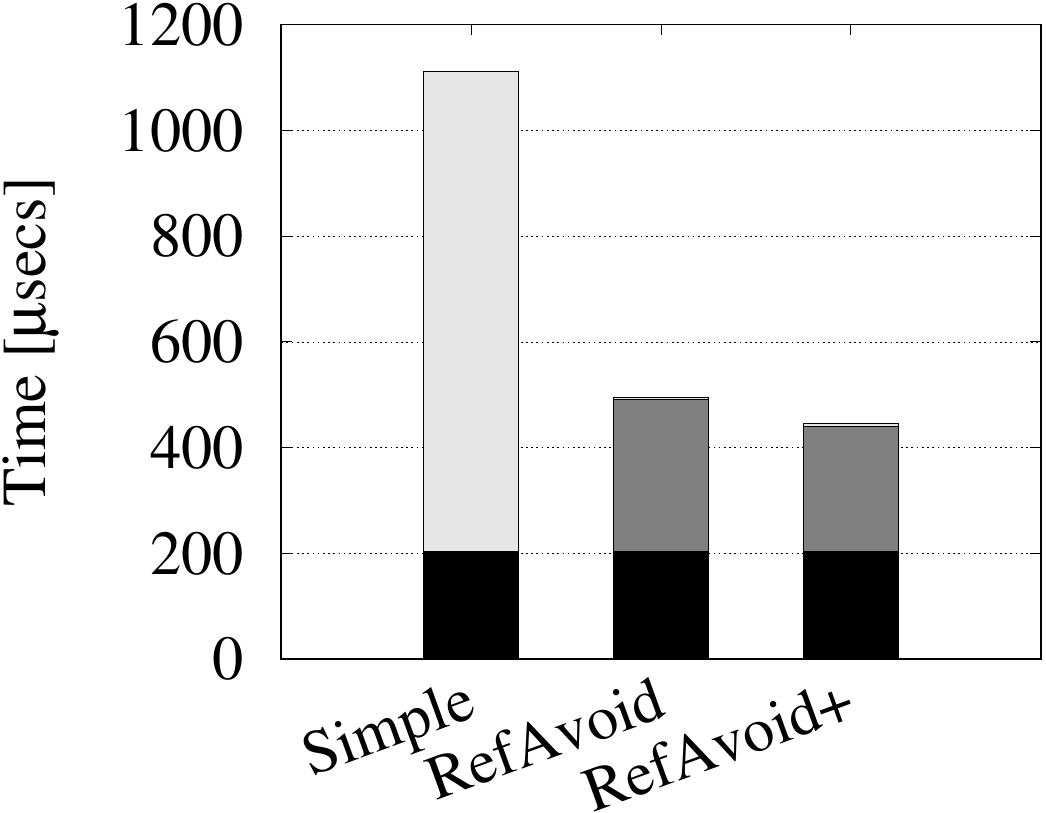}
%{\scriptsize query extent ratio [\%]} &{\scriptsize query extent ratio [\%]} &{\scriptsize query extent ratio [\%]} &{\scriptsize query extent ratio [\%]}
%\end{tabular}
%\caption{Time breakdown on window queries: 2-level with 2000 partitions per dimension, 0.1\% query extent ratio}
%\label{fig:ref}
%\end{figure*}
%\begin{figure*}[t]
%\centering
%\begin{tabular}{cccc}
%AREAWATER &LINEARWATER &ROADS &EDGES\\
\\
\multicolumn{4}{c}{Window queries}\\\\
\includegraphics[width=0.5\columnwidth]{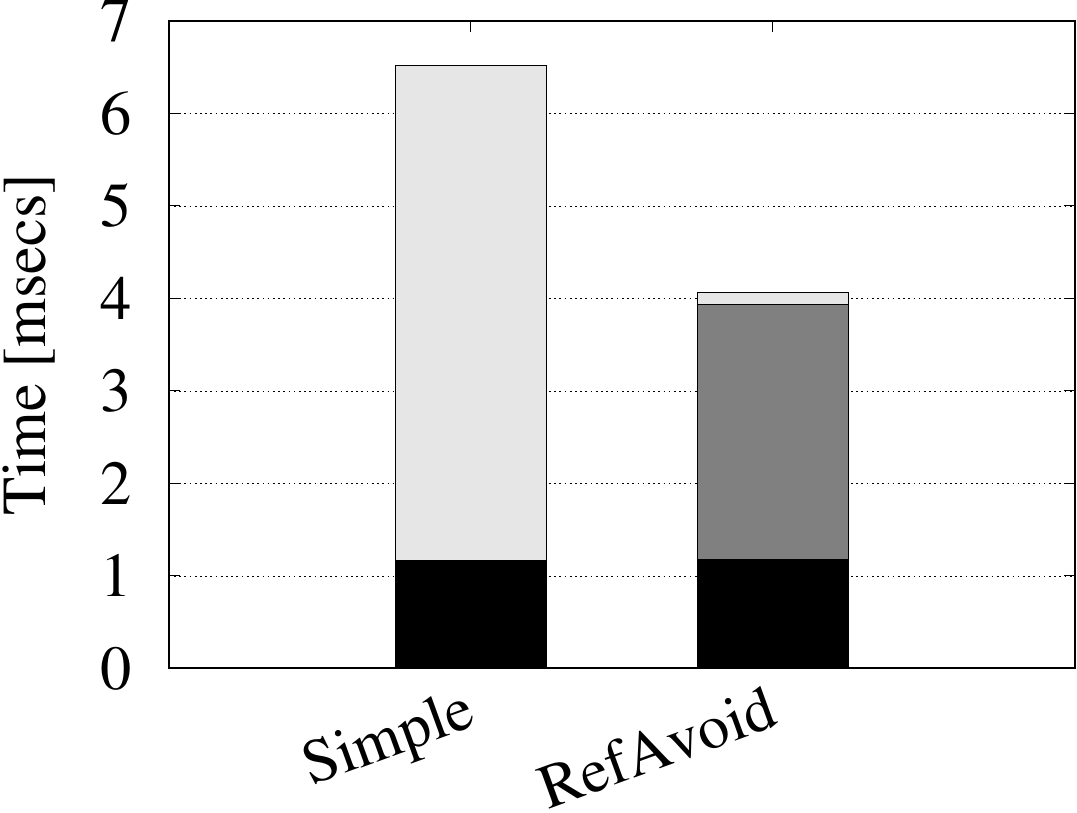}
&\hspace{-1ex}\includegraphics[width=0.5\columnwidth]{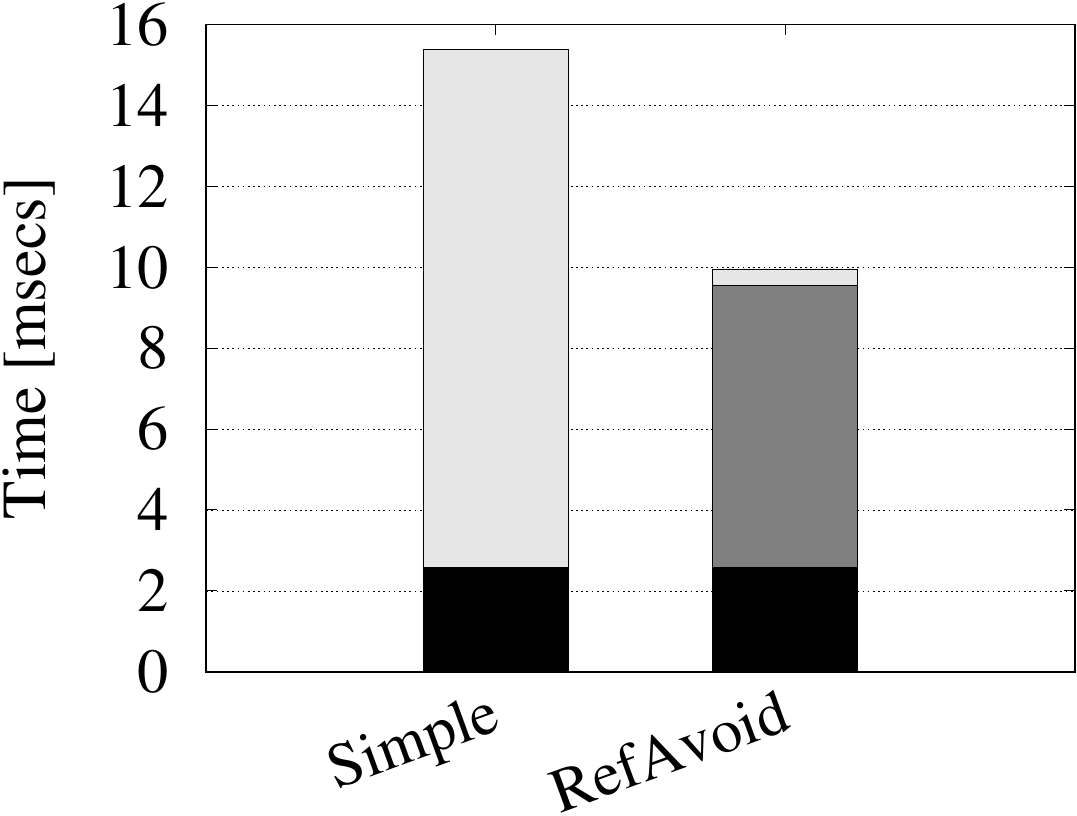}
&\hspace{-1ex}\includegraphics[width=0.5\columnwidth]{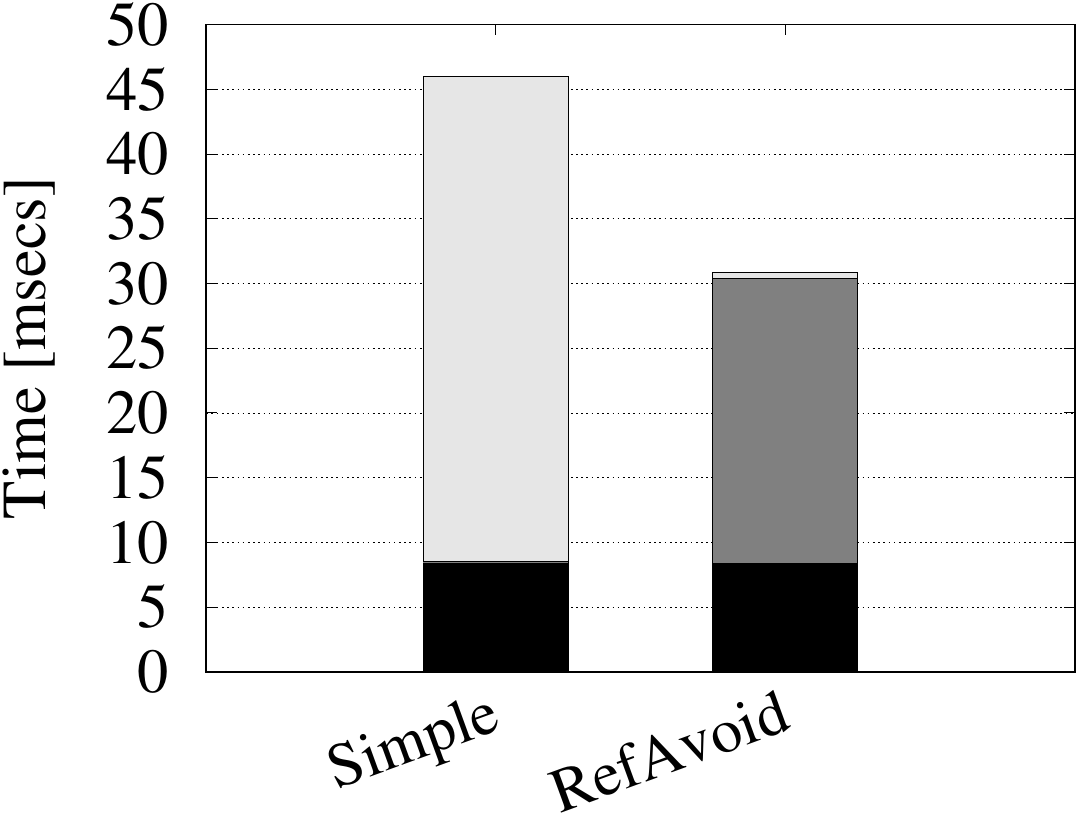}
&\hspace{-1ex}\includegraphics[width=0.5\columnwidth]{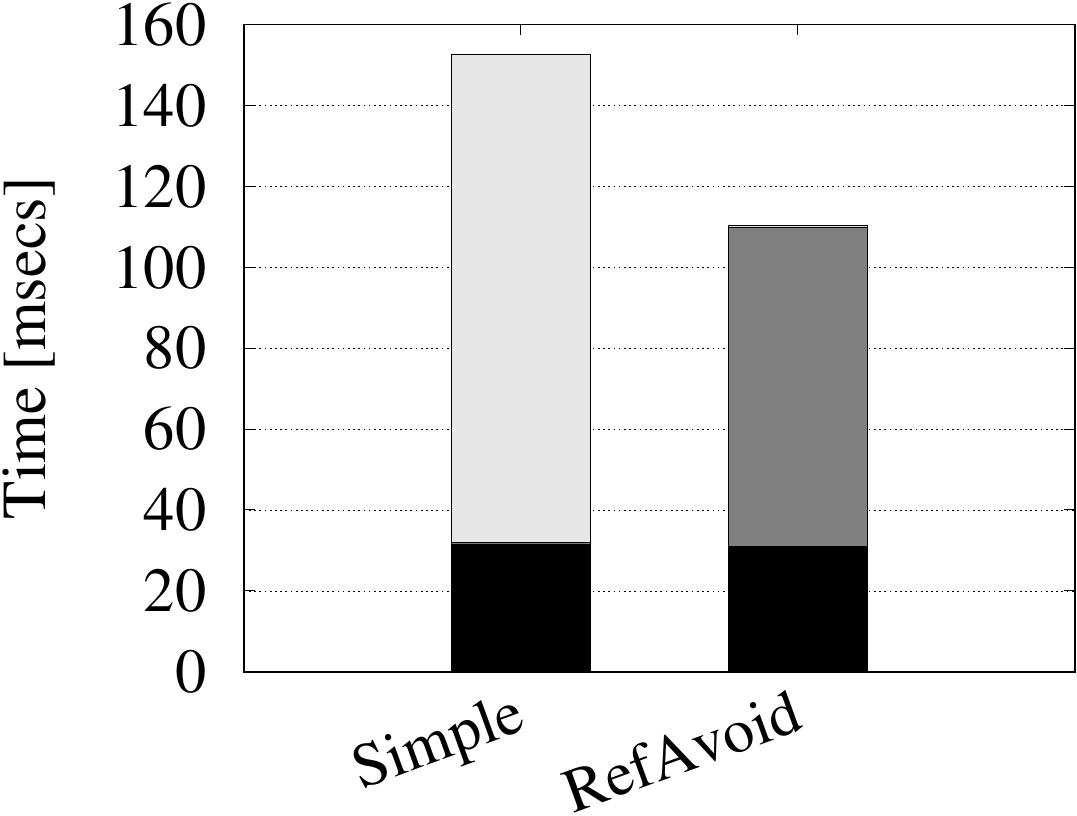}
%{\scriptsize query extent ratio [\%]} &{\scriptsize query extent ratio [\%]} &{\scriptsize query extent ratio [\%]} &{\scriptsize query extent ratio [\%]}
\\
\multicolumn{4}{c}{Disk queries}
\end{tabular}
%\caption{Time breakdown on disc queries: 2-level with 2000 partitions per dimension, 0.1\% query extent ratio}
\caption{Time breakdown: \twolevel indexing with 2000 partitions per dimension, 10000 queries of 0.1\% extent ratio}
\label{fig:ref}
\end{center}
\end{figure*}
We first justify our decision to focus on
and optimize the filtering step of range query evaluation, which in
fact has been the primary target of previous works as well.
We used our \twolevel index to execute both the filtering and
refinement steps.
We consider three variants of query evaluation depending on
the way refinement is performed (see Section~\ref{sec:refine});
filtering is identical in all three variants.
More specifically, under \refvone, all candidates identified by the
filtering step are passed to the refinement step;
\refvtwo employs Lemma~\ref{lemma:ref} as an extra pre-refinement
filter to significantly reduce the number of candidates to be refined;
last, \refvthree enhances \refvtwo by reducing the number of
comparisons required for testing Lemma~\ref{lemma:ref}
in window queries,
as discussed at the end of Section \ref{sec:refine}.
%. \panos{"minimize" is strong wording; it's mentioned also in Section~\ref{sec:refine}. Can we prove that the number of comparison is the minimum or should we just use "reduce"?}

Figure~\ref{fig:ref} illustrates the breakdown of the average
execution time for both window and disk queries;
note that for disk queries \refvthree is not applicable.
We make two important observations.
First, the figure clearly shows the effectiveness of the refinement
avoidance technique discussed in Section~\ref{sec:refine}.
Both \refvtwo and \refvthree significantly reduce the number
of candidates to be refined by over 90\% and so, their refinement step
is always lower than that of \refvone.
To achieve this however, they apply a refinement avoidance test on the
MBRs; the cost of this is higher in case of disk queries because it
requires expensive distance computations between the disk center
and the corners of object MBRs.
The second observation is that,
when our refinement avoidance technique is used,
the bottleneck of the query is in the filtering step.
%not only the queries run faster but also the filtering step (including the pre-refinement filter) is now the most expensive step, instead of refinement in case with \refvone.
Hence, in the subsequent experiments, we focus on the 
%remaining of the section is to show how our two-level index accelerate the
filtering step of spatial query processing.
 
\subsection{Indexing and Tuning}
\begin{figure*}[t]
\centering
\begin{tabular}{cccc}
AREAWATER &LINEARWATER &ROADS &EDGES\\
\includegraphics[width=0.5\columnwidth]{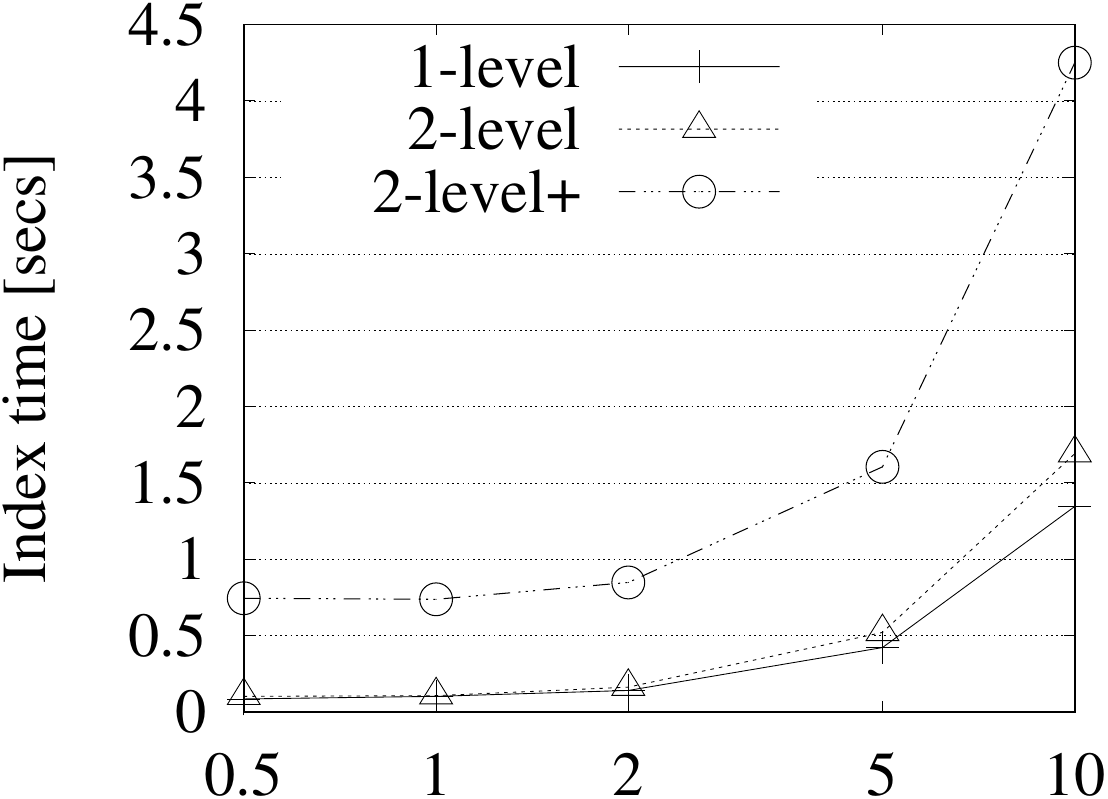}
&\hspace{-1ex}\includegraphics[width=0.5\columnwidth]{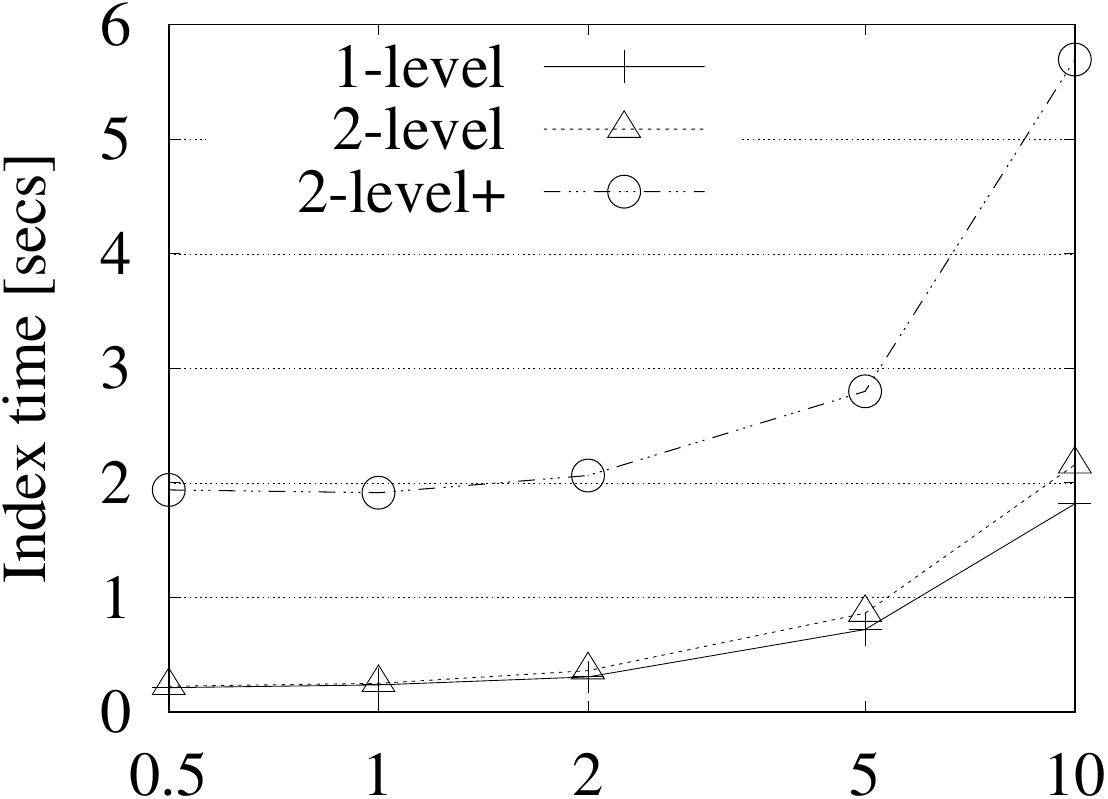}
&\hspace{-1ex}\includegraphics[width=0.5\columnwidth]{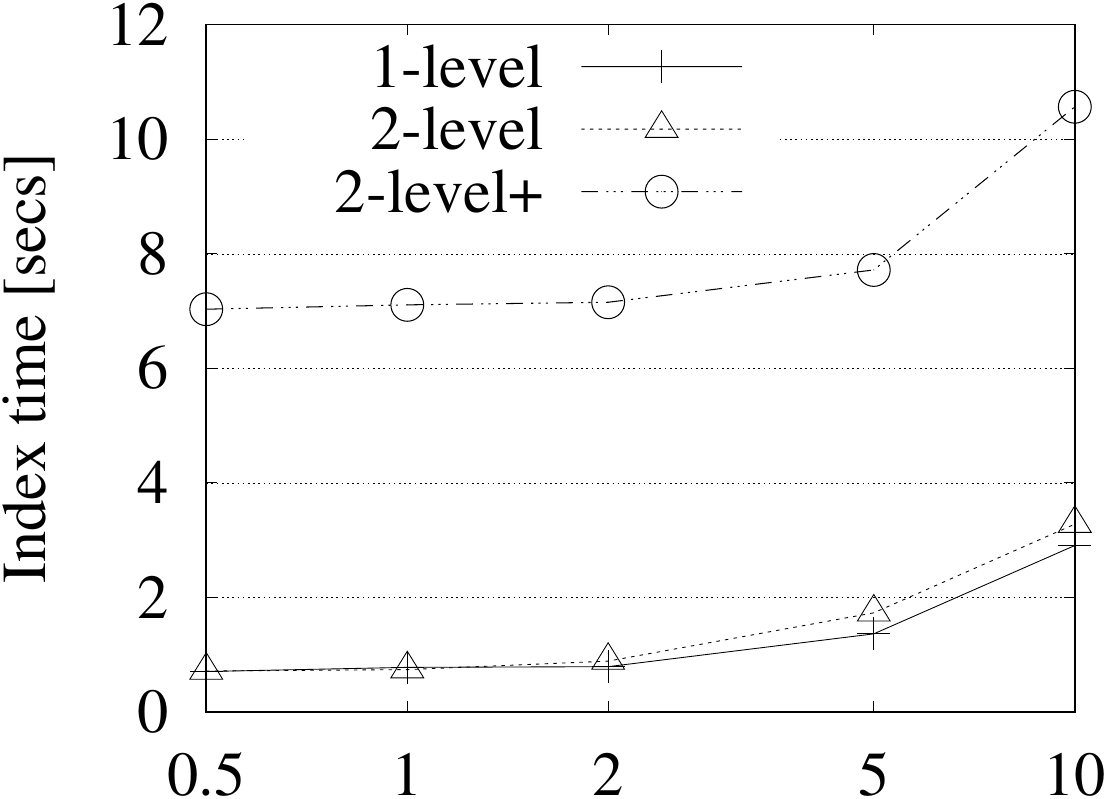}
&\hspace{-1ex}\includegraphics[width=0.5\columnwidth]{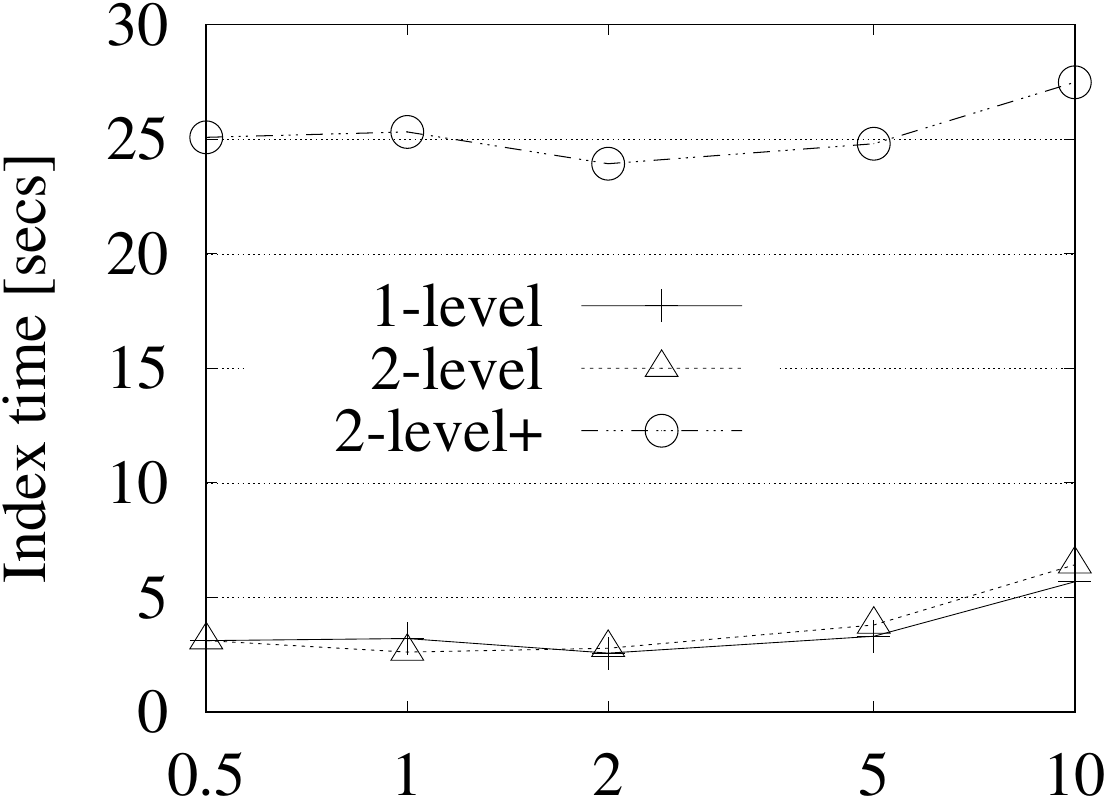}\\
\eat{
\includegraphics[width=0.5\columnwidth]{plots/OLD/qrange_T2_AREAWATER_fixed_mbr_USA_c01_n10000_vary-p_ours_itime.pdf}
&\hspace{-1ex}\includegraphics[width=0.5\columnwidth]{plots/OLD/qrange_T5_LINEARWATER_fixed_mbr_USA_c01_n10000_vary-p_ours_itime.pdf}
&\hspace{-1ex}\includegraphics[width=0.5\columnwidth]{plots/OLD/qrange_T8_ROADS_fixed_mbr_USA_c01_n10000_vary-p_ours_itime.pdf}
&\hspace{-1ex}\includegraphics[width=0.5\columnwidth]{plots/OLD/qrange_T4_EDGES_fixed_mbr_USA_c01_n10000_vary-p_ours_itime.pdf}\\
}
{\scriptsize partitions per dimension [$\times1000$]} &{\scriptsize partitions per dimension [$\times1000$]} &{\scriptsize partitions per dimension [$\times1000$]} &{\scriptsize partitions per dimension [$\times1000$]}\\\\
\includegraphics[width=0.5\columnwidth]{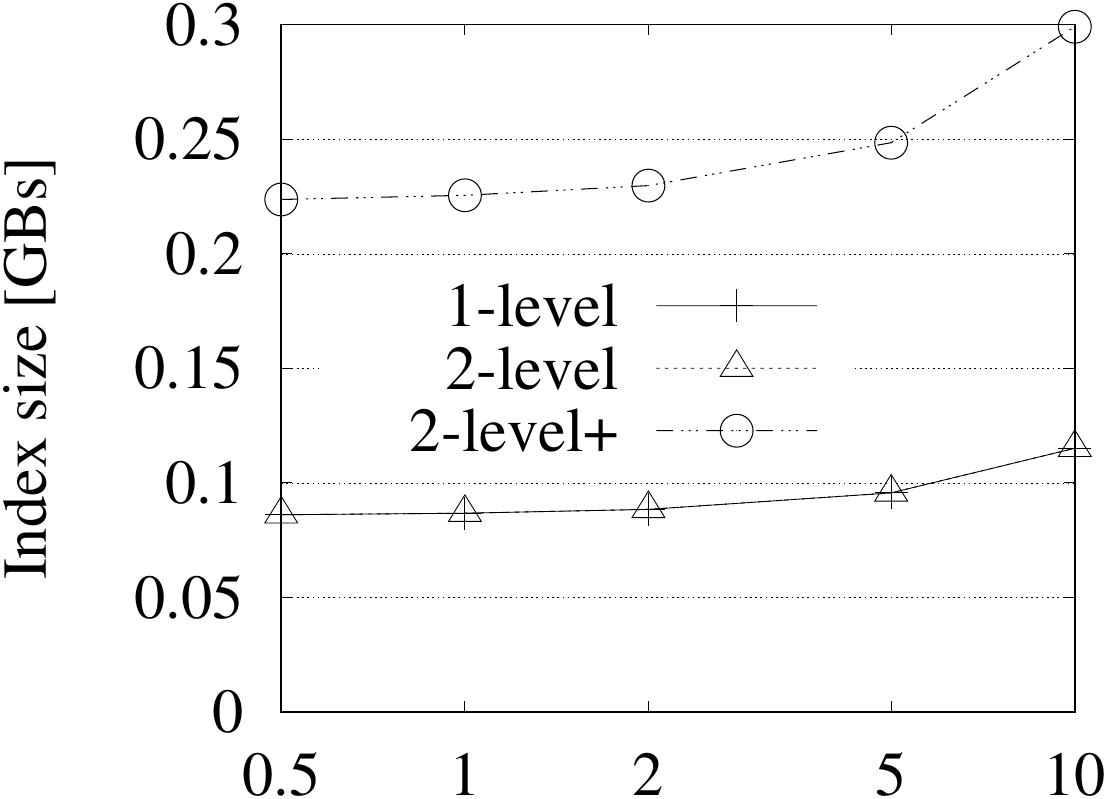}
&\hspace{-1ex}\includegraphics[width=0.5\columnwidth]{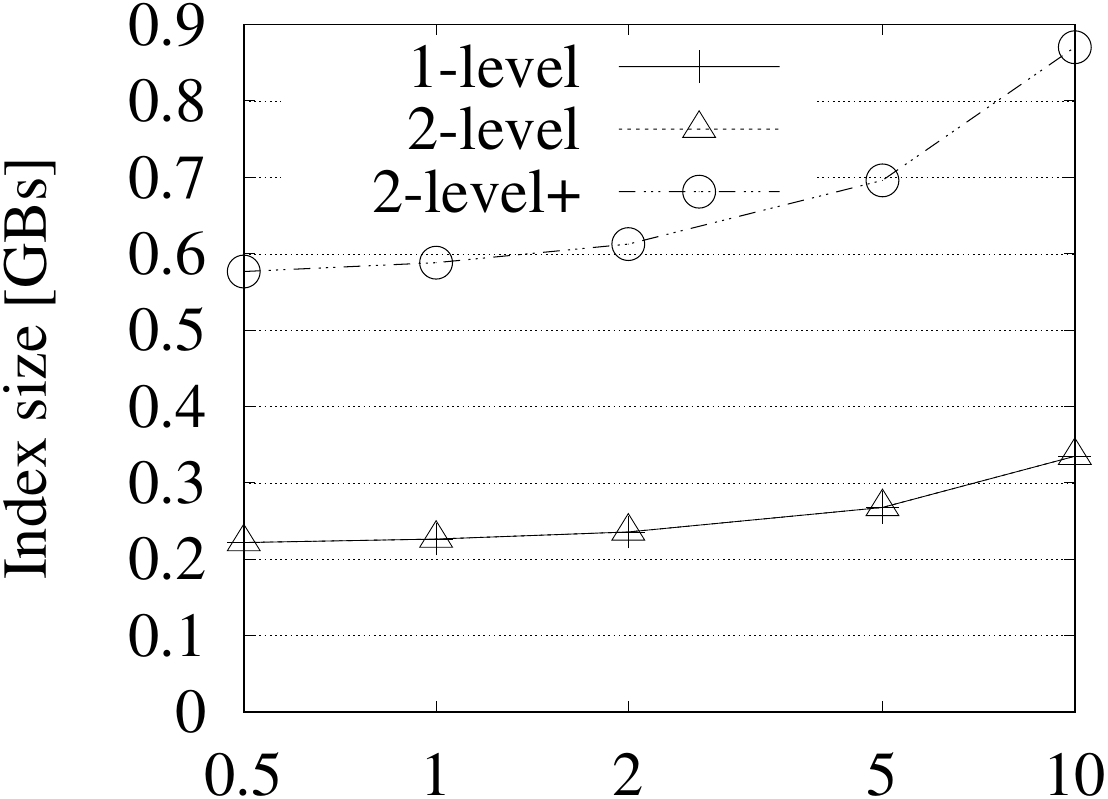}
&\hspace{-1ex}\includegraphics[width=0.5\columnwidth]{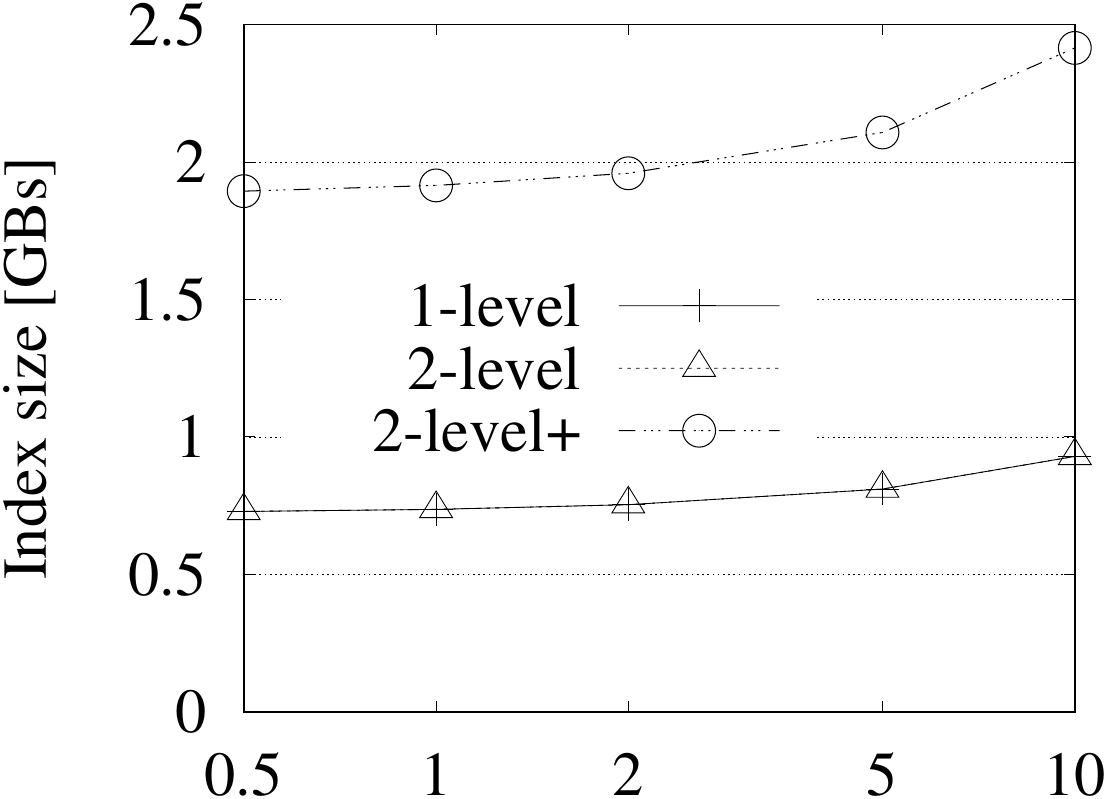}
&\hspace{-1ex}\includegraphics[width=0.5\columnwidth]{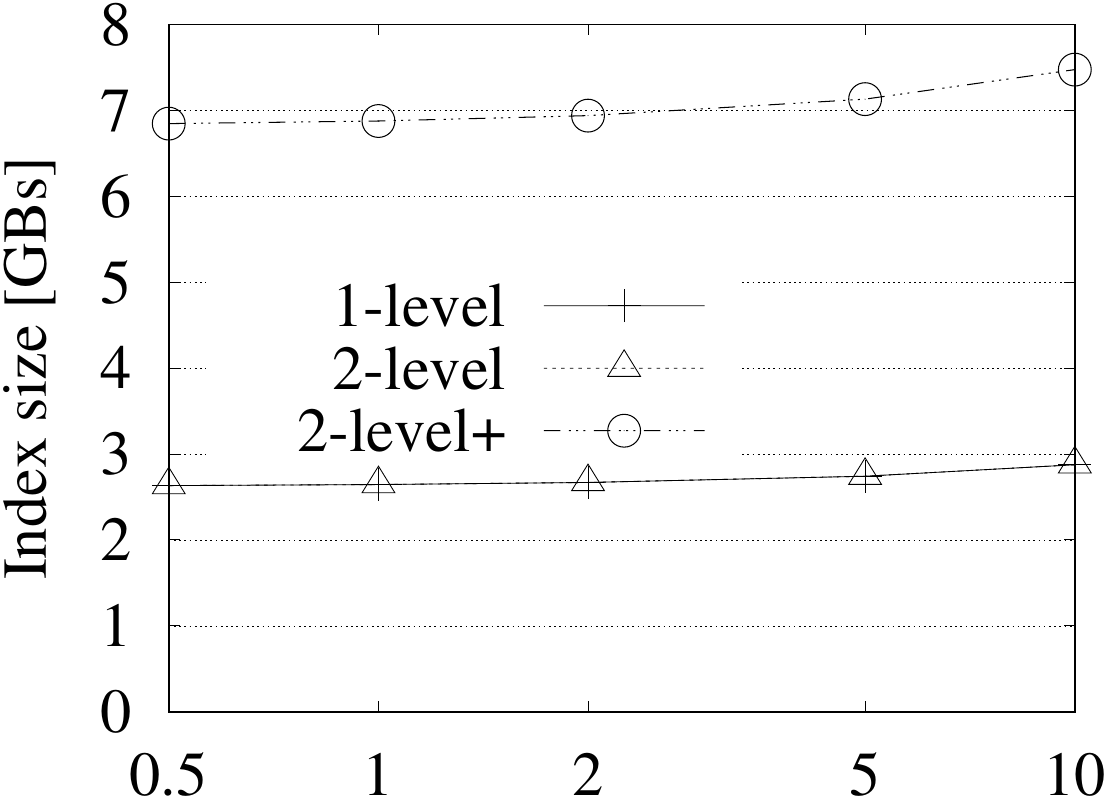}\\
{\scriptsize partitions per dimension [$\times1000$]} &{\scriptsize partitions per dimension [$\times1000$]} &{\scriptsize partitions per dimension [$\times1000$]} &{\scriptsize partitions per dimension [$\times1000$]}\\\\
\includegraphics[width=0.5\columnwidth]{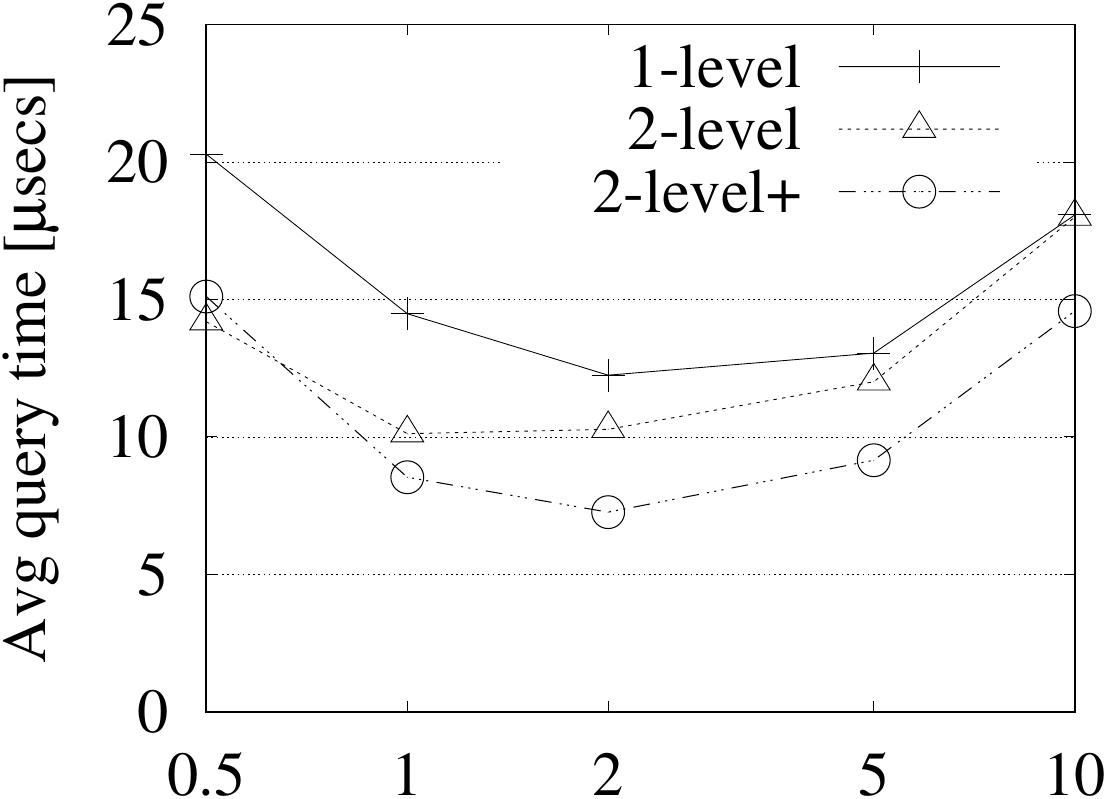}
&\hspace{-1ex}\includegraphics[width=0.5\columnwidth]{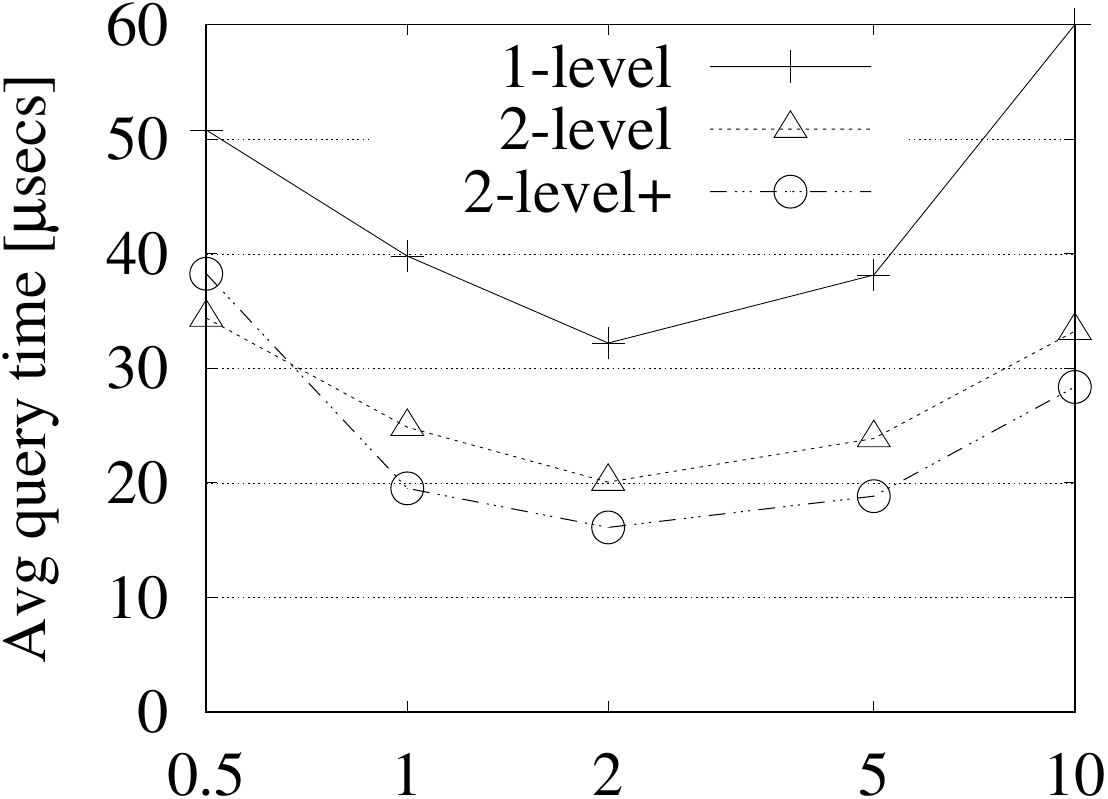}
&\hspace{-1ex}\includegraphics[width=0.5\columnwidth]{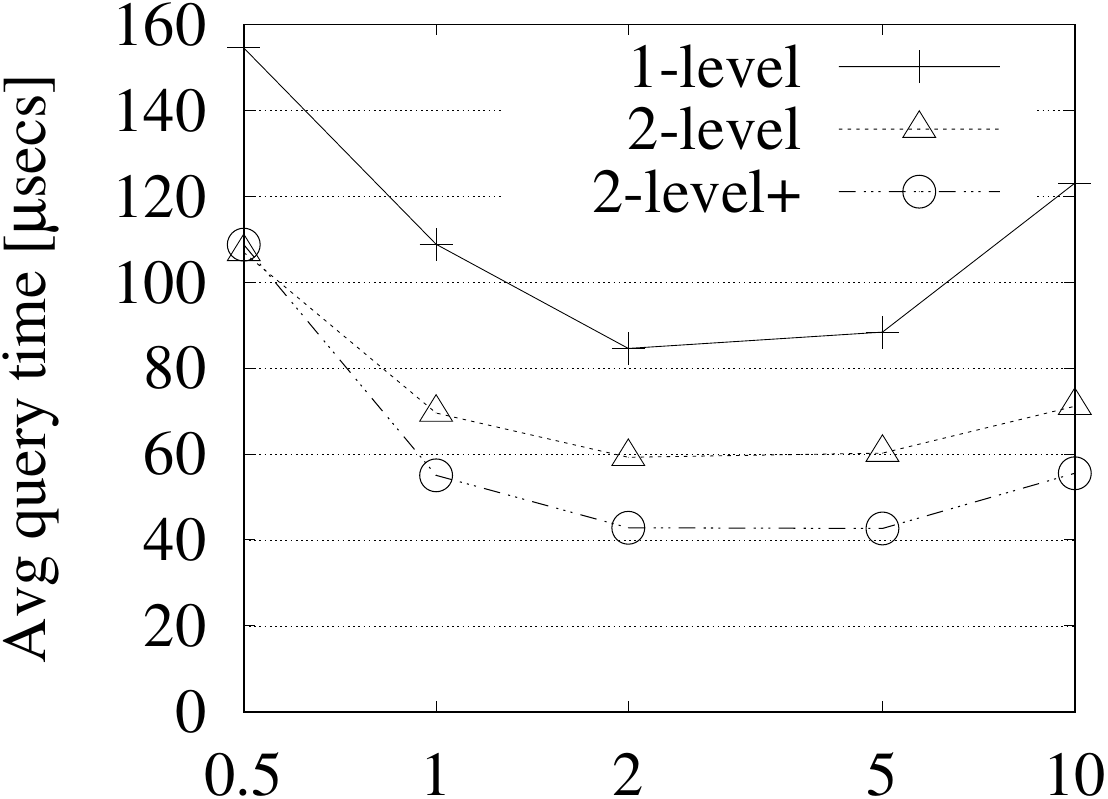}
&\hspace{-1ex}\includegraphics[width=0.5\columnwidth]{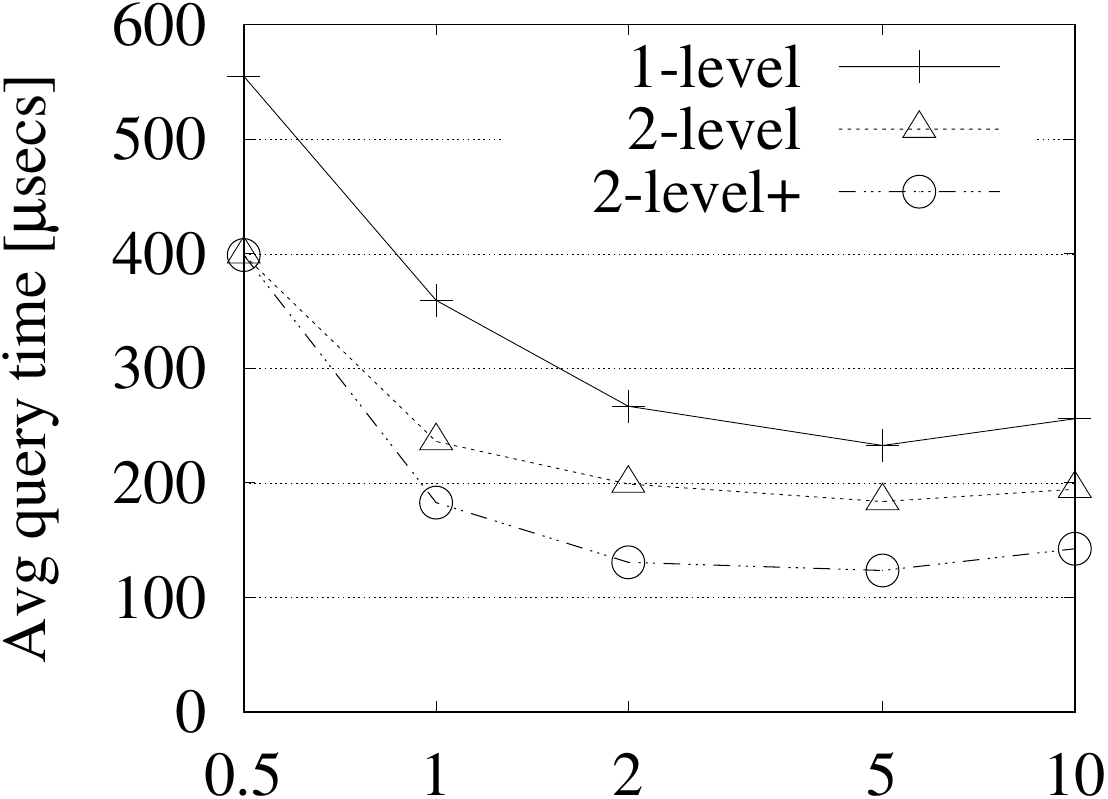}\\
{\scriptsize partitions per dimension [$\times1000$]} &{\scriptsize partitions per dimension [$\times1000$]} &{\scriptsize partitions per dimension [$\times1000$]} &{\scriptsize partitions per dimension [$\times1000$]}
\eat{\\
\includegraphics[width=0.5\columnwidth]{plots/OLD_beforeCodeOptimizations/qrange_T2_AREAWATER_fixed_mbr_USA_c01_n10000_vary-p_ours.pdf}
&\hspace{-1ex}\includegraphics[width=0.5\columnwidth]{plots/OLD_beforeCodeOptimizations/qrange_T5_LINEARWATER_fixed_mbr_USA_c01_n10000_vary-p_ours.pdf}
&\hspace{-1ex}\includegraphics[width=0.5\columnwidth]{plots/OLD_beforeCodeOptimizations/qrange_T8_ROADS_fixed_mbr_USA_c01_n10000_vary-p_ours.pdf}
&\hspace{-1ex}\includegraphics[width=0.5\columnwidth]{plots/OLD_beforeCodeOptimizations/qrange_T4_EDGES_fixed_mbr_USA_c01_n10000_vary-p_ours.pdf}\\
\includegraphics[width=0.5\columnwidth]{plots/OLD2/qrange_T2_AREAWATER_fixed_mbr_USA_c01_n10000_vary-p_ours.pdf}
&\hspace{-1ex}\includegraphics[width=0.5\columnwidth]{plots/OLD2/qrange_T5_LINEARWATER_fixed_mbr_USA_c01_n10000_vary-p_ours.pdf}
&\hspace{-1ex}\includegraphics[width=0.5\columnwidth]{plots/OLD2/qrange_T8_ROADS_fixed_mbr_USA_c01_n10000_vary-p_ours.pdf}
&\hspace{-1ex}\includegraphics[width=0.5\columnwidth]{plots/OLD2/qrange_T4_EDGES_fixed_mbr_USA_c01_n10000_vary-p_ours.pdf}\\
\includegraphics[width=0.5\columnwidth]{plots/OLD/qrange_T2_AREAWATER_fixed_mbr_USA_c01_n10000_vary-p_ours.pdf}
&\hspace{-1ex}\includegraphics[width=0.5\columnwidth]{plots/OLD/qrange_T5_LINEARWATER_fixed_mbr_USA_c01_n10000_vary-p_ours.pdf}
&\hspace{-1ex}\includegraphics[width=0.5\columnwidth]{plots/OLD/qrange_T8_ROADS_fixed_mbr_USA_c01_n10000_vary-p_ours.pdf}
&\hspace{-1ex}\includegraphics[width=0.5\columnwidth]{plots/OLD/qrange_T4_EDGES_fixed_mbr_USA_c01_n10000_vary-p_ours.pdf}\\
}
\end{tabular}
\caption{Indexing and tuning: varying the granularity of the grid, 10000 window queries of 0.1\% relative extent}
\label{fig:vary-p}
\end{figure*}
We next investigate the building cost and the tuning of our two-level
index. Figure~\ref{fig:vary-p} reports the indexing time for both
\twolevel and \twolevelplus variants and the size of each index, while
varying the granularity of the underlying grid partitioning. For
reference and completeness purposes, we include the \onelevel
competitor which also uses a uniform grid to partition the input
space. As the goal of the index is to efficiently answer spatial range
queries, we additionally report the average query time in order to
determine the best grid granularity.

Naturally, the indexing cost for all three indices grows while
increasing the granularity of the grid.
%However, there are important differences.
As expected, both \onelevel and \twolevel have the same space
requirements.
Regardless of employing the second level of partitioning or not,
both indices store exactly the same number of object MBRs (originals
and replicas);
the difference is that inside each tile, \twolevel divides the
rectangles in four classes and stores them in
dedicated structures while \onelevel stores all rectangles together.
In terms of the indexing time, \twolevel is slightly more expensive
than \onelevel,
as it needs to first determine the class for each rectangle
and then store it accordingly.
On the other hand, the indexing cost of \twolevelplus is higher than
both \onelevel and \twolevel indices.
Remember that \twolevelplus essentially stores a second copy of the
rectangles inside every tile,
after decomposing their coordinate information. As
a result, the size of \twolevelplus is 2.4x larger; its
building time is also higher due to the cost of computing,
sorting and storing the decomposed replicas of the rectangles.
The sizes of the packed R-trees (not shown) are about the same as the
sizes of the corresponding \onelevel (and \twolevel) indices when 2000
partitions per dimension are used, indicating that the replication
ratio of our indexes is very low.
In addition, the bulk loading costs of the R-trees are 0.53s, 1.4s,
5.2s, and 19.5s for the four datasets, respectively, i.e., about 20\%
lower compared to the construction cost of \twolevelplus. 
%, i.e., the number of partitions per dimension. 

Let us now study the efficiency of each index. The key observation
is that employing the second level of indexing significantly
enhances query processing; \twolevel and \twolevelplus always
outperform \onelevel. Due to lack of space, Figure~\ref{fig:vary-p}
reports only on window queries, but the same trend applies for disk
queries.
The fastest index is \twolevelplus as it trades the extra used space
for better query performance; nevertheless, \twolevel is also
significantly faster than \onelevel. The last observation
is that all three indices perform at their best when the
underlying grid defines approximately 2000 partitions per dimension.
Under this configuration, the number of tiles is not excessive
and the indices do not have a large overhead in accessing and managing tiles;
at the same time the rectangles are small enough compared
to the tile extent (see Table \ref{tab:datasets}), in order not to
incur excessive replication.
%\fix{Justify/confirm this using the statistics!}.
For the rest of our analysis, \onelevel, \twolevel and
\twolevelplus
all use a 2000$\times$2000 uniform grid.

\subsection{Cost of Updates}
In order to confirm the superiority of our proposed indexing scheme in
updates compared to the
R-tree, we conducted an experiment, where for all datasets, we first
constructed the index by loading 90\% of the data in batch and then
we measured the cost of incrementally inserting the last 10\% of the data.
Table \ref{tab:updates_cost} compares the total update costs of the
R-tree, \onelevel, and \twolevel.
As the table shows, the R-tree is two orders of magnitude slower than
the baseline \onelevel index and the cost of updates on \twolevel is
only a bit higher compared to the update cost on \onelevel.

\eat{
% 95/5
\begin{table}
\centering
\caption{Total update cost (msec)}
\label{tab:updates_cost}
\begin{tabular}{|c|c|c|c|}
\hline
\textbf{dataset}		&\textsf{R-tree}	&\onelevel		&\twolevel\\\hline\hline
%AREAWATER			&0.307991			&0.00422435	&0.00501077\\
AREAWATER			&308					&4.2 				&5.1\\
%LINERWATER		&0.823199			&0.0117089	&0.0133655\\
LINERWATER		&823						&11.7					&13.3\\
%ROADS					&2.89321			&0.0305563	&0.0340345\\
ROADS					&2893					&31					&34\\
%EDGES					&1.01927							&0.00982038	&0.0113302\\
EDGES					&1019					&9.8					&11.3\\
\hline
\end{tabular}
\end{table}
}
% 90/10
\begin{table}
\centering
\caption{Total update cost (sec)}
\label{tab:updates_cost}
\begin{tabular}{|c|c|c|c|}
\hline
\textbf{dataset}		&\textsf{R-tree}	&\onelevel		&\twolevel\\\hline\hline
AREAWATER			&0.619			&0.007	&0.009\\
LINEARWATER		&1.574			&0.023	&0.027\\
ROADS		&5.34		&0.059	&0.068\\
EDGES					&19.8							&0.220	&0.241538\\
\hline
\end{tabular}
\end{table}

\subsection{Querying Processing}
\begin{figure*}[t]
\centering
\begin{tabular}{cccc}
AREAWATER &LINEARWATER &ROADS &EDGES\\
\includegraphics[width=0.5\columnwidth]{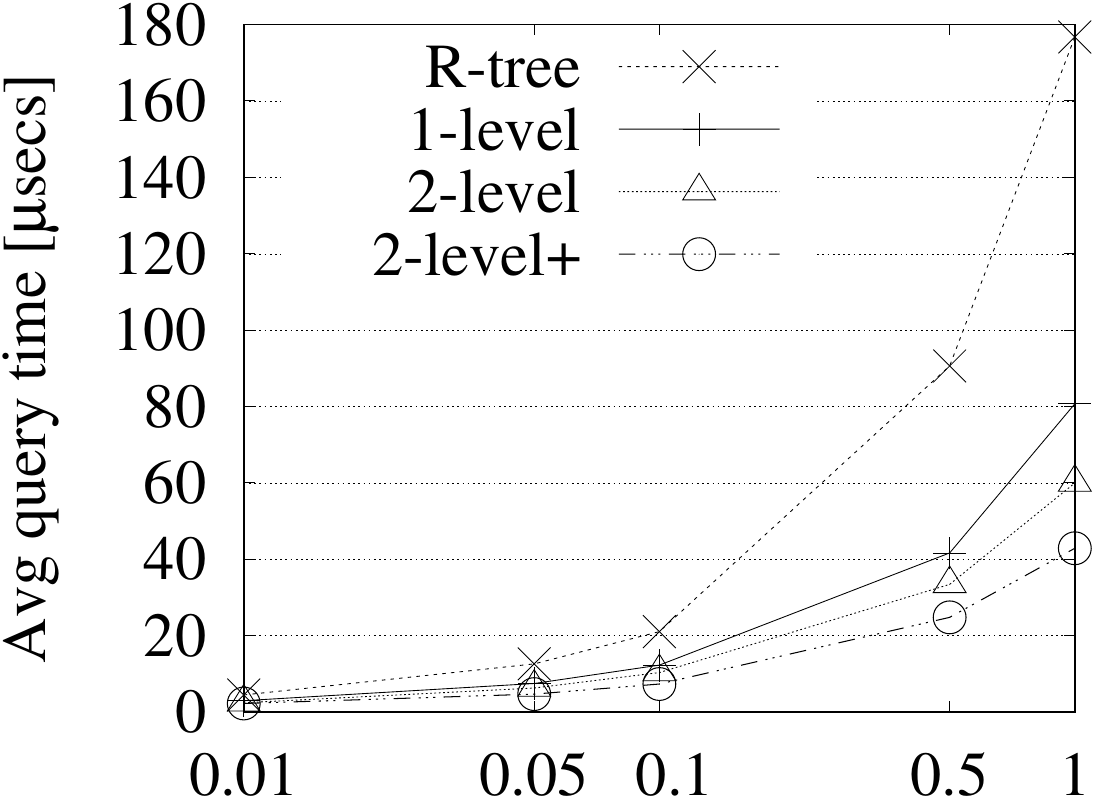}
&\hspace{-1ex}\includegraphics[width=0.5\columnwidth]{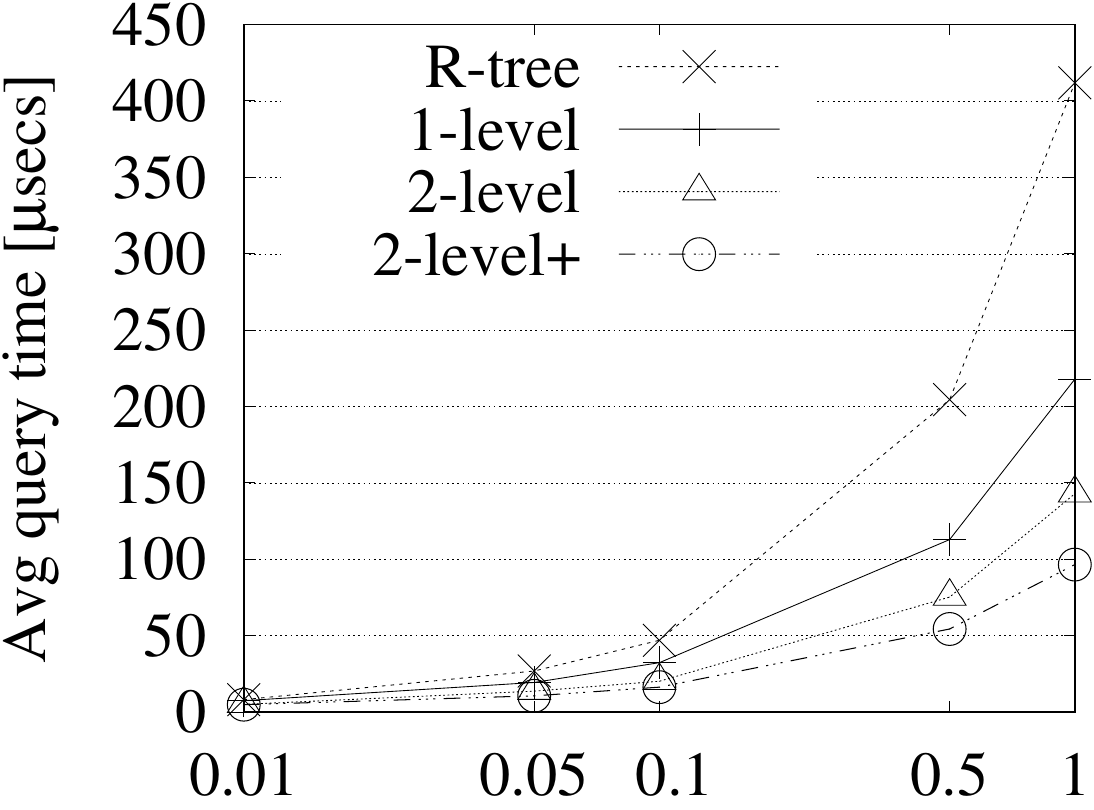}
&\hspace{-1ex}\includegraphics[width=0.5\columnwidth]{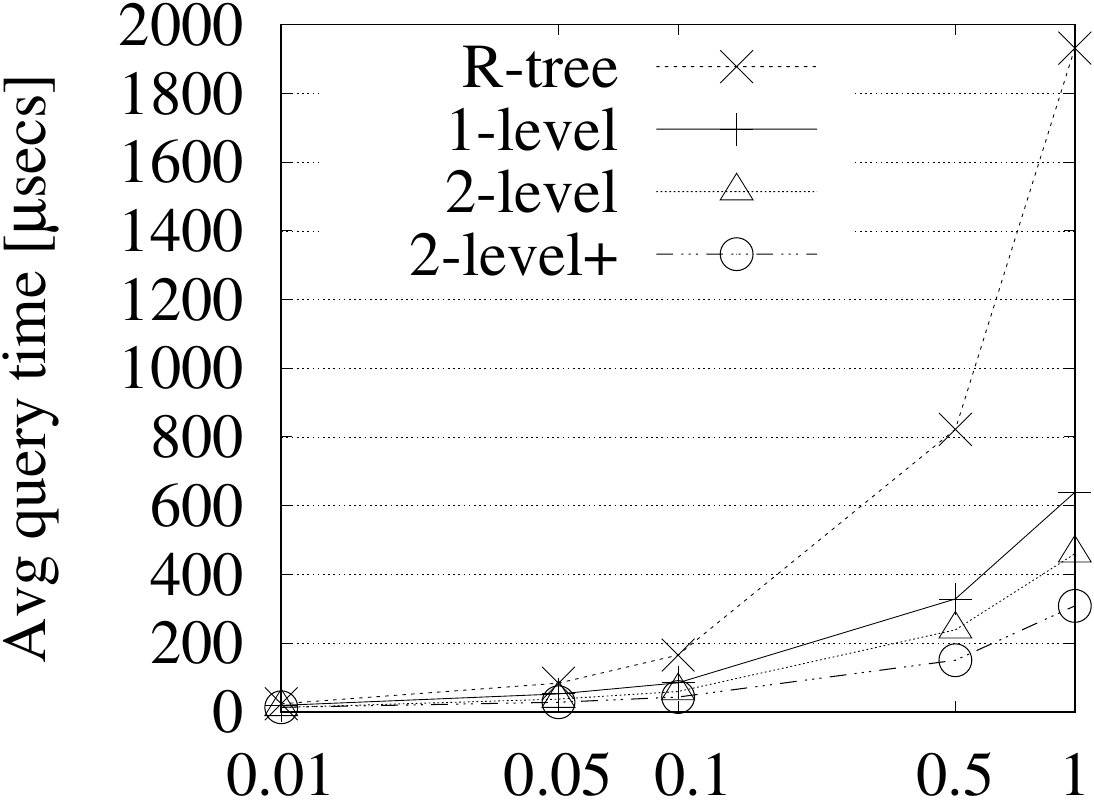}
&\hspace{-1ex}\includegraphics[width=0.5\columnwidth]{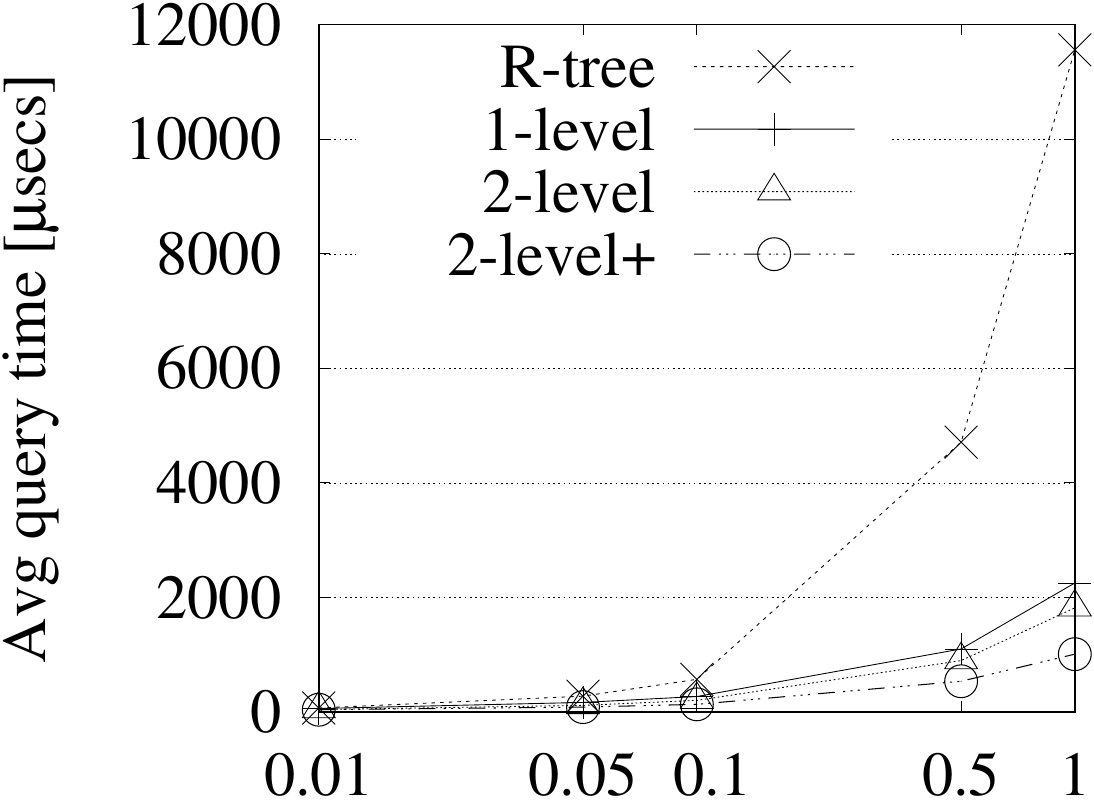}\\
{\scriptsize query relative extent [\%]} &{\scriptsize query relative extent [\%]} &{\scriptsize query relative extent [\%]} &{\scriptsize query relative extent [\%]}
\eat{\\
\includegraphics[width=0.5\columnwidth]{plots/OLD_beforeCodeOptimizations/qrange_T2_AREAWATER_fixed_mbr_USA_n10000_vary-c_comps.pdf}
&\hspace{-1ex}\includegraphics[width=0.5\columnwidth]{plots/OLD_beforeCodeOptimizations/qrange_T5_LINEARWATER_fixed_mbr_USA_n10000_vary-c_comps.pdf}
&\hspace{-1ex}\includegraphics[width=0.5\columnwidth]{plots/OLD_beforeCodeOptimizations/qrange_T8_ROADS_fixed_mbr_USA_n10000_vary-c_comps.pdf}
&\hspace{-1ex}\includegraphics[width=0.5\columnwidth]{plots/OLD_beforeCodeOptimizations/qrange_T4_EDGES_fixed_mbr_USA_n10000_vary-c_comps.pdf}\\
\includegraphics[width=0.5\columnwidth]{plots/OLD/qrange_T2_AREAWATER_fixed_mbr_USA_n10000_vary-c_comps.pdf}
&\hspace{-1ex}\includegraphics[width=0.5\columnwidth]{plots/OLD/qrange_T5_LINEARWATER_fixed_mbr_USA_n10000_vary-c_comps.pdf}
&\hspace{-1ex}\includegraphics[width=0.5\columnwidth]{plots/OLD/qrange_T8_ROADS_fixed_mbr_USA_n10000_vary-c_comps.pdf}
&\hspace{-1ex}\includegraphics[width=0.5\columnwidth]{plots/OLD/qrange_T4_EDGES_fixed_mbr_USA_n10000_vary-c_comps.pdf}\\
}
\\
\multicolumn{4}{c}{Window queries}\\\\
%\end{tabular}
%\caption{Window queries: varying query extent over entire space, 2000 partitions per dimension for 1-level, 2-level and 2-level+}
%\label{fig:vary-c}
%\end{figure*}
%\begin{figure*}[t]
%\centering
%\begin{tabular}{cccc}
%AREAWATER &LINEARWATER &ROADS &EDGES\\
\includegraphics[width=0.5\columnwidth]{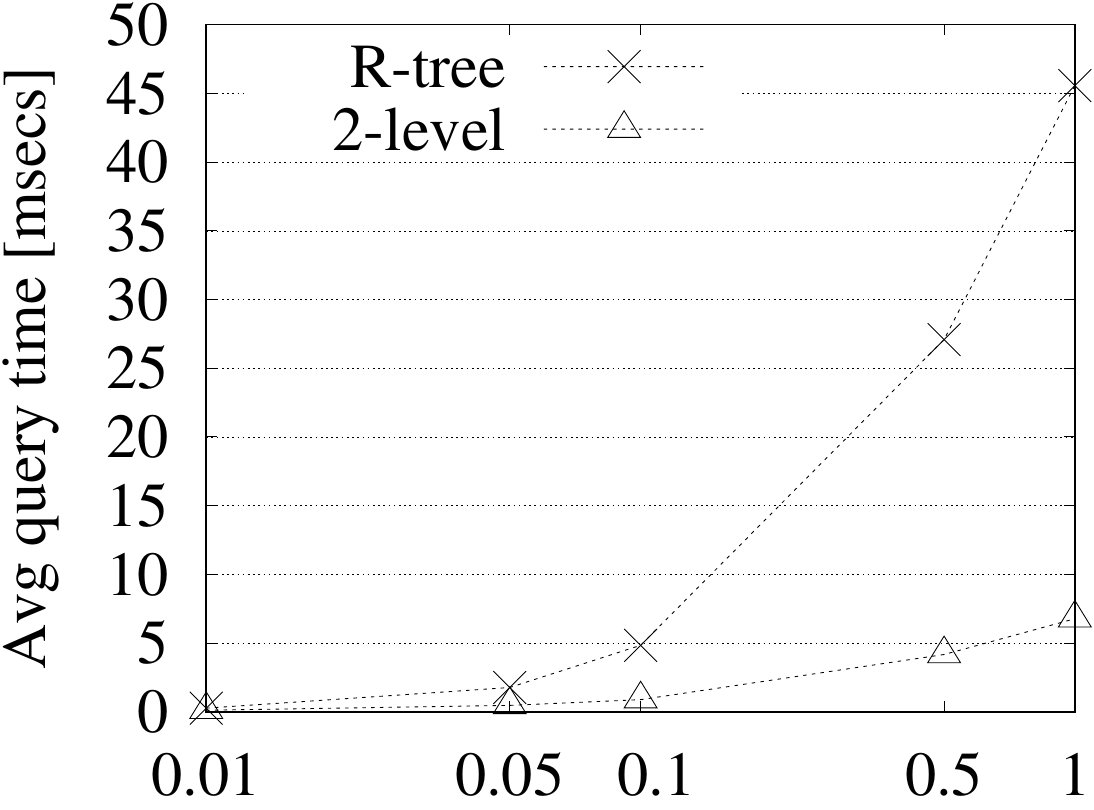}
&\hspace{-1ex}\includegraphics[width=0.5\columnwidth]{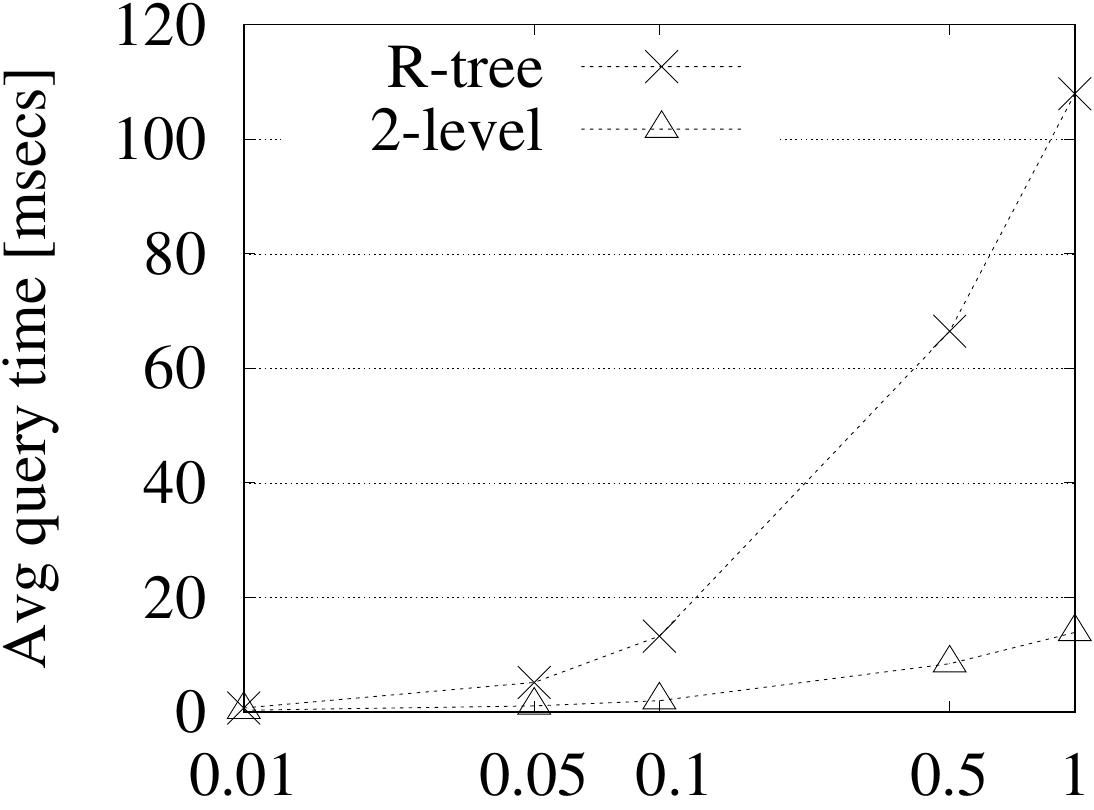}
&\hspace{-1ex}\includegraphics[width=0.5\columnwidth]{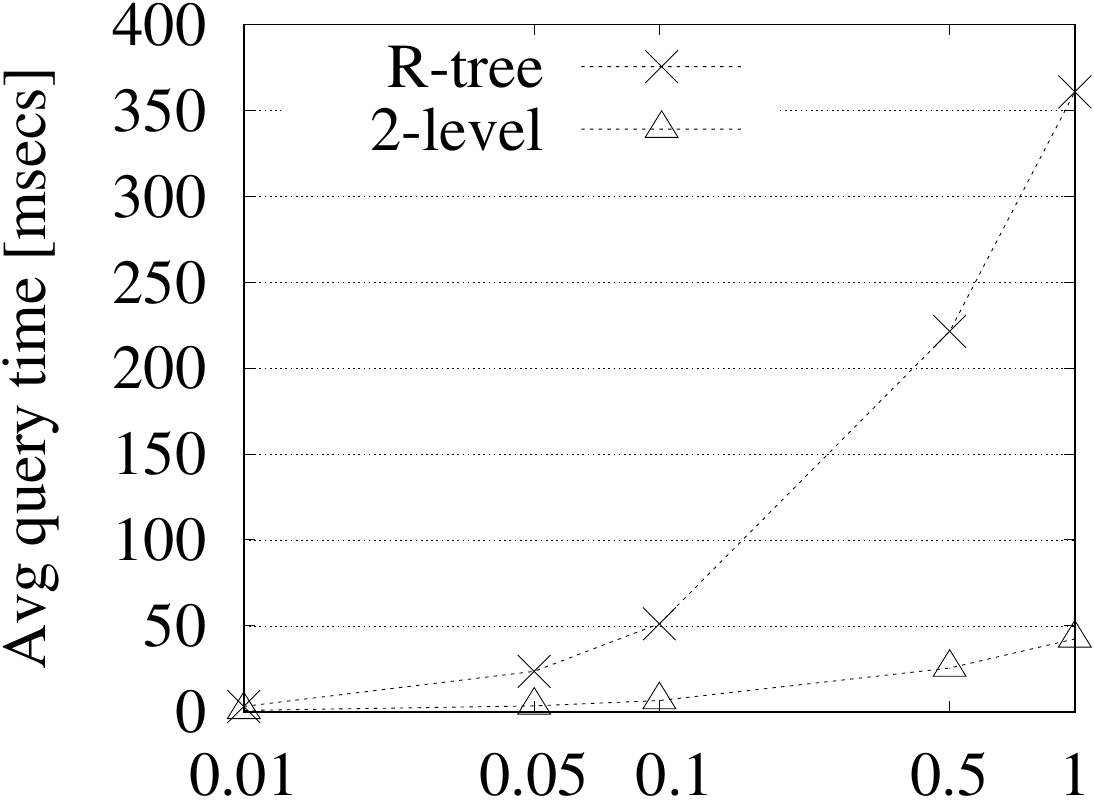}
&\hspace{-1ex}\includegraphics[width=0.5\columnwidth]{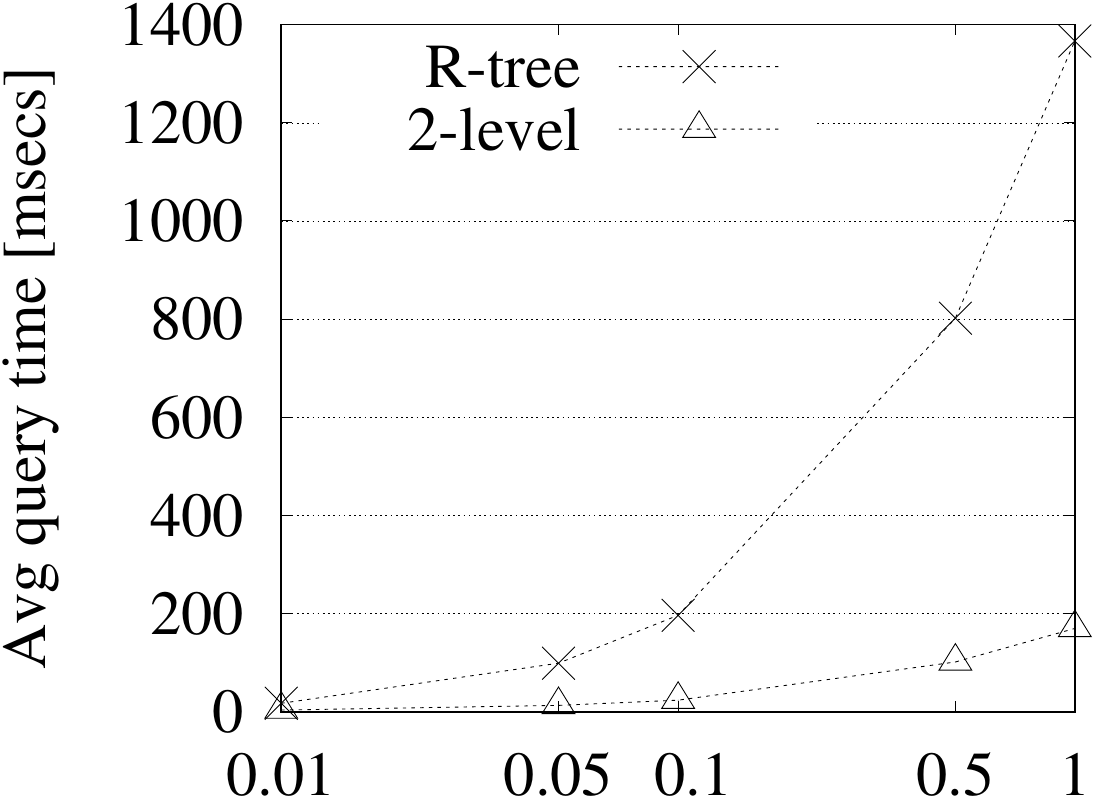}\\
{\scriptsize query relative extent [\%]} &{\scriptsize query relative extent [\%]} &{\scriptsize query relative extent [\%]} &{\scriptsize query relative extent [\%]}\\
\multicolumn{4}{c}{Disk queries}
\end{tabular}
\caption{Query processing: varying query relative extent, 2000 partitions per dimension for \onelevel, \twolevel, \twolevelplus}
\label{fig:vary-c}
\end{figure*}
We now evaluate the performance of the indexes in query processing.
We first compare them in terms of their average query cost and then evaluate batch and parallel query processing.

\stitle{Window queries}.
The first row of plots in Figure~\ref{fig:vary-c} reports the average
execution for window queries, while varying the relative area
of the window compared to the data space.
%\fix{explain how queries are generated. how many queries? and that you  compute  the average.}
The plots also include the performance of the R-tree.
Naturally, query processing is negatively affected by the increase of the relative window extent;
as the windows grow larger they overlap a larger number of objects, rendering the range queries more expensive.
Regarding the comparison between \onelevel, \twolevel and \twolevelplus, we observe the same trend as Figure~\ref{fig:vary-p}.
More importantly though, we observe that our two-level index variants clearly outperform also the R-tree, in all tests, because (i)
the relevant partitions to each query are found very fast (without the need of traversing a hierarchical index) and (ii)
they manage to drastically reduce the total number of computations.
As expected, query processing using \twolevelplus is the most efficient method for window queries and
\twolevel comes in second place.

\stitle{Disk range queries}.
We turn our focus now to disk range queries and the second row of
plots in Figure~\ref{fig:vary-c}.
Different to window queries, the \twolevelplus index does not give any
benefit compared to \twolevel,
because all coordinates of the object MBRs are needed to compute their
distances to the center of the disk (i.e., the decomposed tables are
not useful in computations).
Hence, \twolevelplus is not included in the comparison (as it uses the
complete MBR table in each class and has the
same performance as \twolevel).
In addition, \onelevel we cannot use the reference point technique to
eliminate duplicate results;
instead all tiles use a shared hash table to insert all query results
in order to eliminate duplicates.
This explodes the cost of \onelevel (it becomes one order of magnitude
slower than \twolevel), so we skip it from the comparisons. 
   %    hashing which renders the \onelevel-based computation impractical. Under this, the plots report the average query for R-tree and \twolevel based evaluation.
The results clearly show once again the superiority of the \twolevel index.
The advantage of \twolevel over the R-tree is more pronounced compared
to the case of window queries, as \twolevel manages to avoid the distance
computations  for the majority of query results (which belong to tiles covered by the query range).
%. \fix{+++ intuition why}
%It is also important to observe the difference in time between R-tree and \twolevel in case of the disk queries compared to window. Due to the expensive distance computations required for disk queries, \fix{+++ blah blah, why}, the advantage of \twolevel over the R-tree is more pronounced compared to the case of window queries.

\begin{figure*}[t]
\centering
\begin{tabular}{cccc}
AREAWATER &LINEARWATER &ROADS &EDGES\\
\includegraphics[width=0.5\columnwidth]{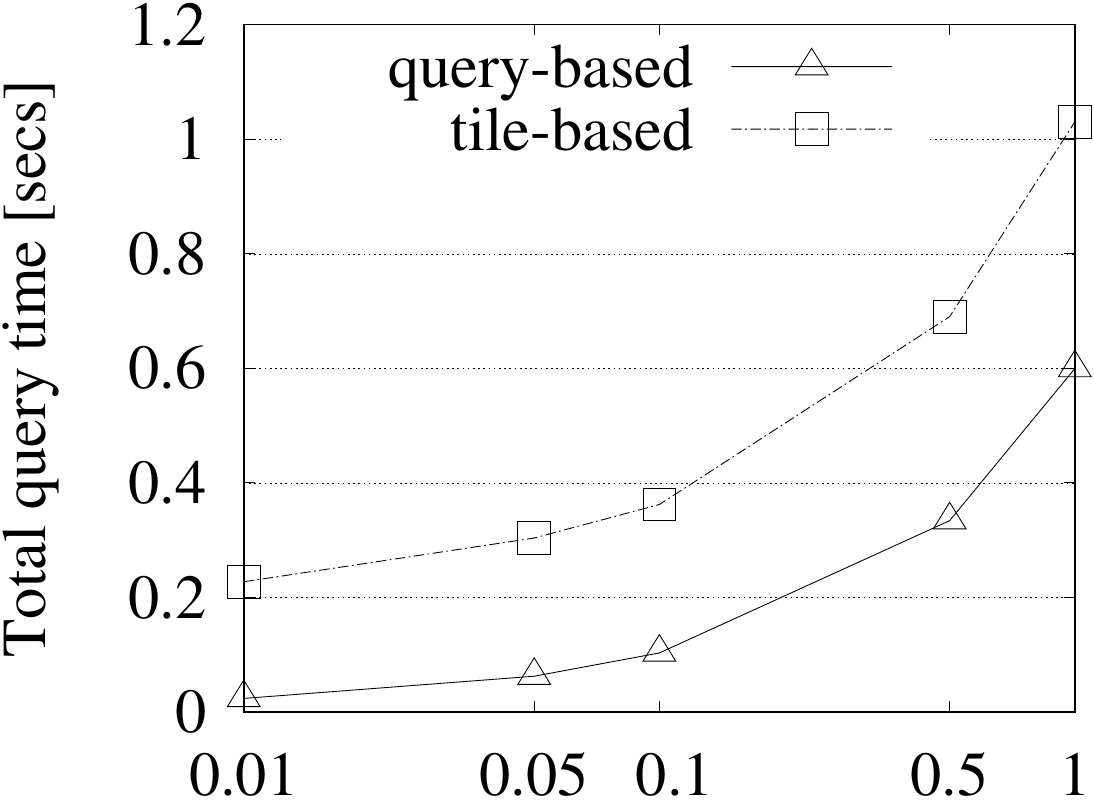}
&\hspace{-1ex}\includegraphics[width=0.5\columnwidth]{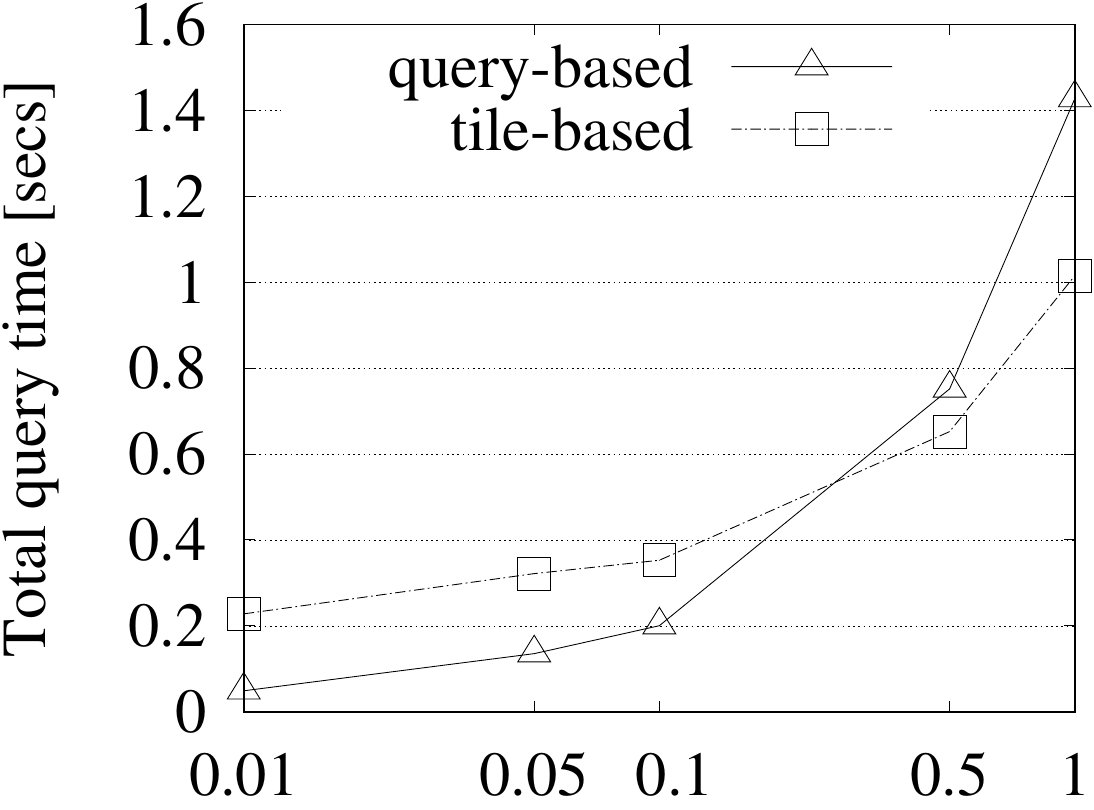}
&\hspace{-1ex}\includegraphics[width=0.5\columnwidth]{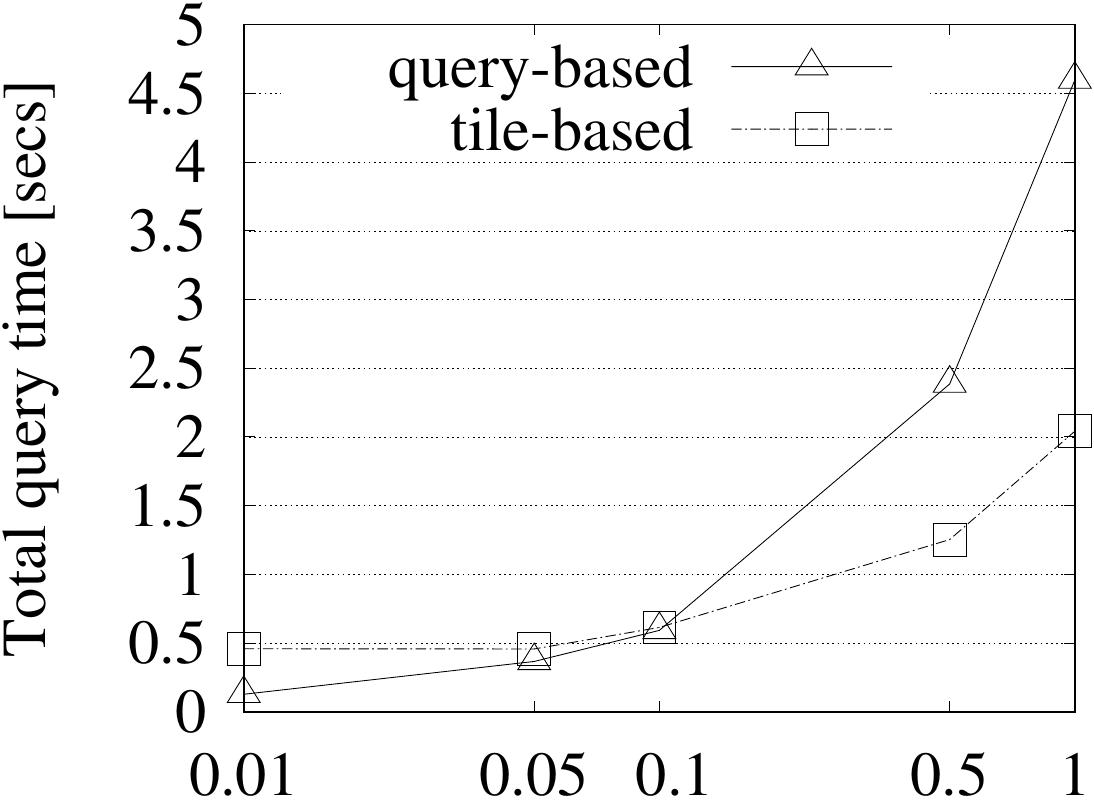}
&\hspace{-1ex}\includegraphics[width=0.5\columnwidth]{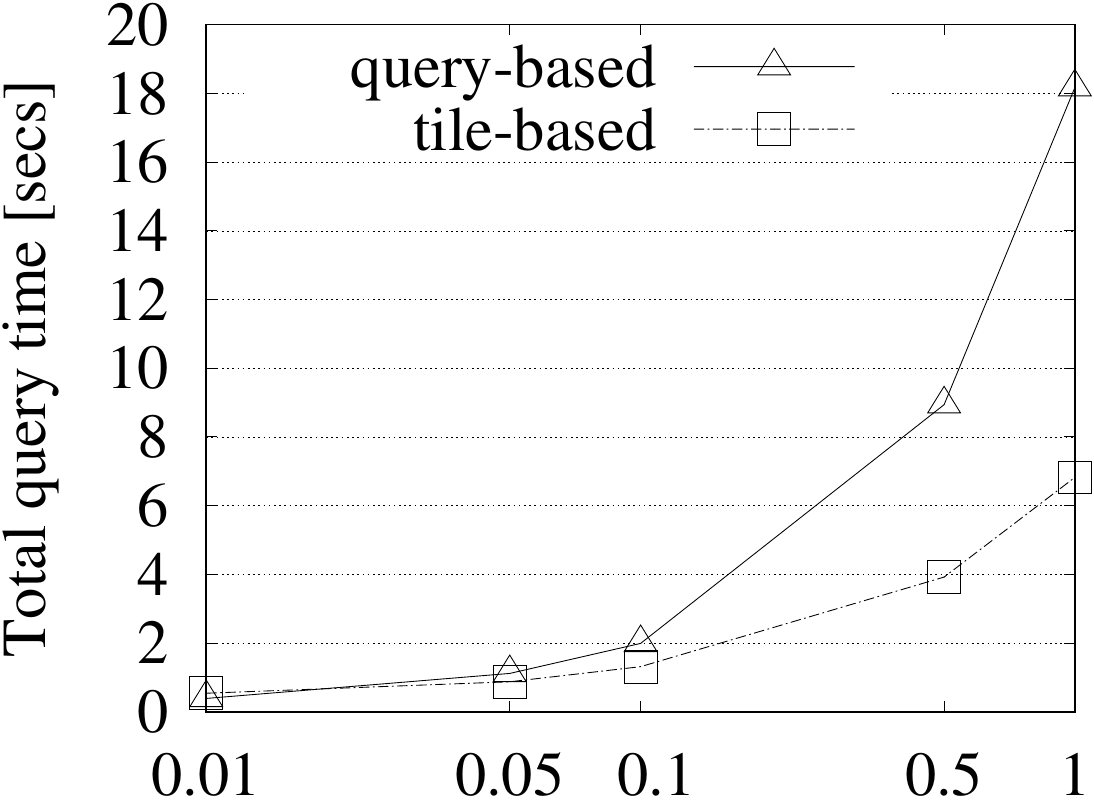}\\
{\scriptsize query relative extent [\%]} &{\scriptsize query relative
                                           extent [\%]} &{\scriptsize
                                                          query
                                                          relative
                                                          extent [\%]}
                              &{\scriptsize query relative extent
                                [\%]}\\
%\\
%TODO &TODO &TODO &TODO\\
%{\scriptsize \# queries [$\times$1000]} &{\scriptsize \# queries [$\times$1000]} &{\scriptsize \# queries [$\times$1000]} &{\scriptsize \# queries [$\times$1000]}
\end{tabular}
\caption{Batch query processing for window queries: 2000 partitions per dimension}
\label{fig:vary-c_batch}
\end{figure*}

\begin{figure}[t]
\centering
\begin{tabular}{cc}
ROADS &EDGES\\
\hspace{-1ex}\includegraphics[width=0.48\columnwidth]{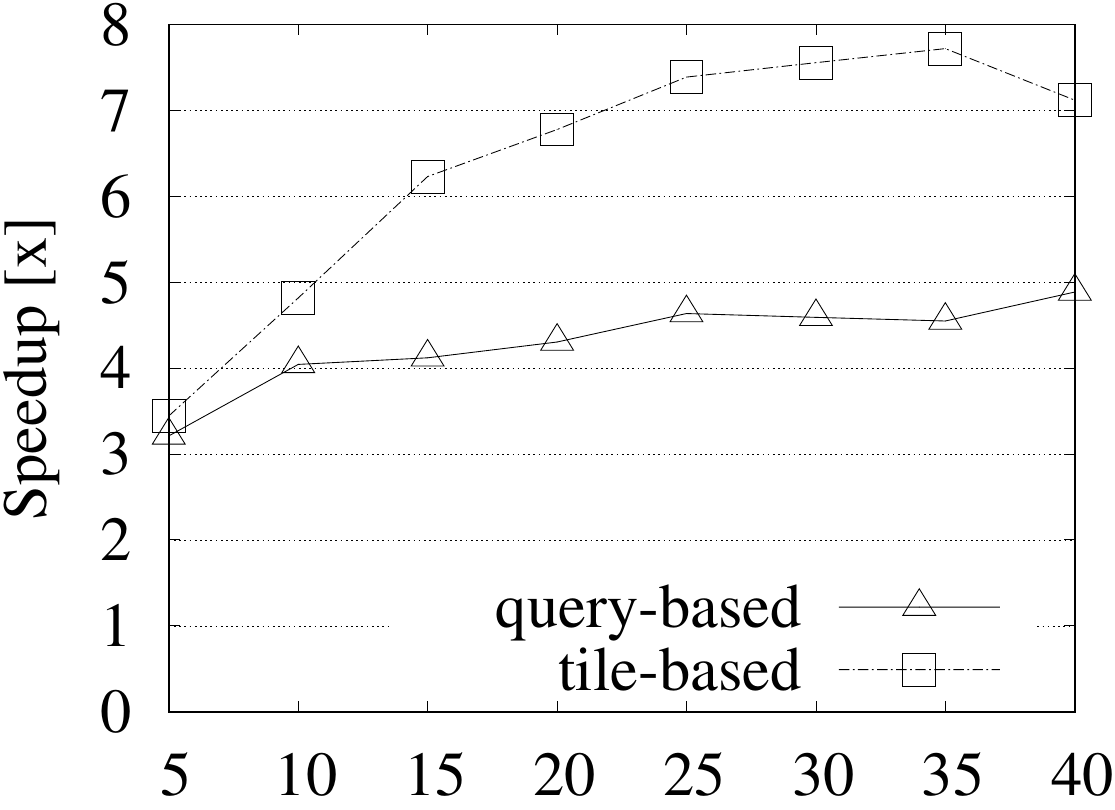}
&\hspace{-1ex}\includegraphics[width=0.48\columnwidth]{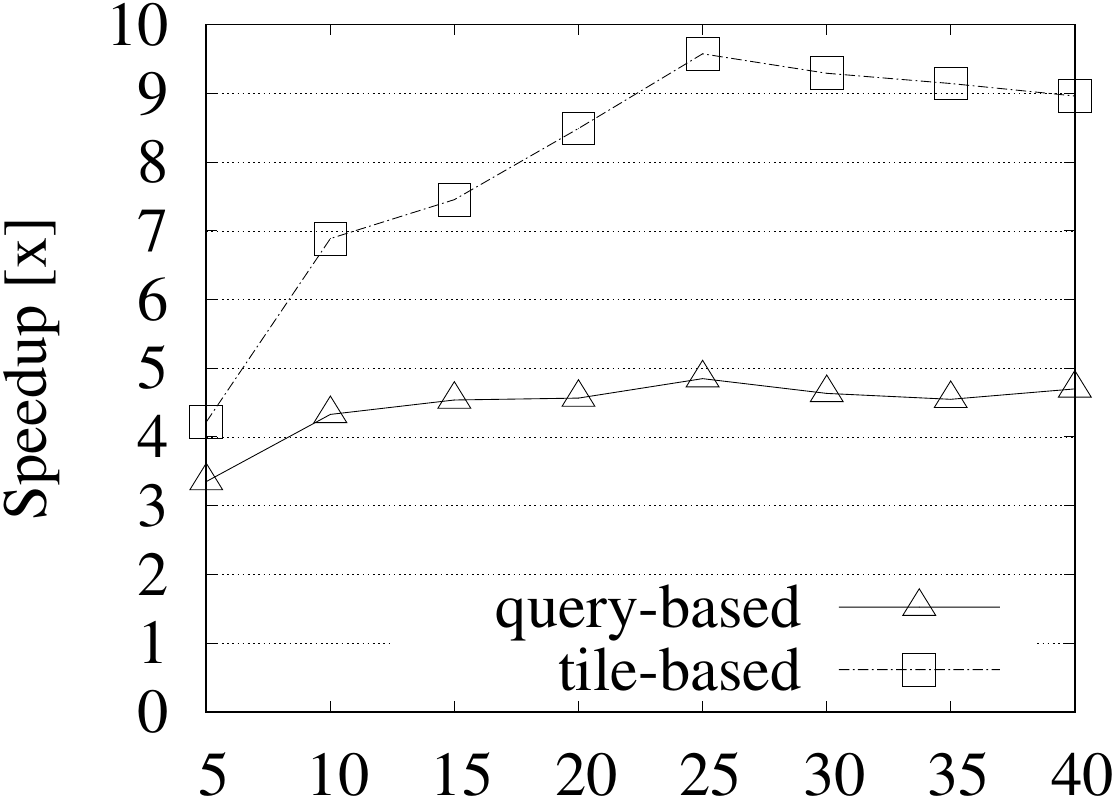}\\
{\scriptsize \# threads} &{\scriptsize \# threads}
\end{tabular}
\caption{Batch query parallel processing: 10000 window queries of 1\% relative extent, 2000 partitions per dimension}
\label{fig:vary-c_batch_parallel}
\end{figure}

\stitle{Batch and parallel query processing}.
Figure \ref{fig:vary-c_batch} compares the two approaches (\qatomic
and \tatomic), discussed in Section \ref{sec:parallel}, for batch window query processing
(10K queries per batch).
%single-threading processing of window query batches (10K queries each).
A general observation from the plots is that \tatomic is superior to 
\qatomic when the dataset is large (i.e., dense) and the queries are
relatively large. In this case, the sizes of the dedicated tables for
each class per tile are large and cache conscious  \tatomic approach
makes a difference. On the other hand, the overhead of finding and accumulating
the subtasks per tile does not pay off when the number of queries
on each tile is too small or when the tiles do not contain many rectangles.
%\panos{write me}
% cheap queries, small datasets, small batches => do queries-atomic
The advantage of \tatomic becomes more prominent in parallel query
processing.  Figure \ref{fig:vary-c_batch_parallel}
shows the speedup of batch query evaluation
on the two largest datasets
(again, 10K queries per
batch)
as a function of the
number of parallel threads.
Note that  \tatomic scales gracefully with the number of threads (up
to about 25 threads, where it starts being affected by
hyperthreading).
On the other hand, \qatomic scales poorly due to the numerous cache misses.

%\subsection{Batch Query Processing} 

%

%cheap queries, small datasets, small batches => do queries-atomic

%Parallel processing: due to lack of space only window queries, blah blah

\section{Conclusions}
\label{sec:conclusion}
In this paper, we presented a secondary partitioning approach that can
be applied to space-partitioning spatial indexes, such as grids and
divides the indexed rectangles within each spatial partition (tile) to
four classes. Our approach reduces the number of comparisons during
range query evaluation and avoids the generation (and elimination) of
duplicate results. In addition, we propose a refinement avoidance
technique for spatial range queries, which confirms as results the
great majority of objects that intersect the range without needing to
apply a refinement step for them. Finally, we investigate techniques
for evaluating numerous range query requests in batch and in parallel.
Our experimental findings confirm the efficiency of our proposed
indexing scheme compared to a state-of-the-art in-memory
implementation of the R-tree and its scalability to multiple query
evaluation in parallel.

In the future, we will dig more into parallel and distributed spatial
query evaluation. The fact that our indexing scheme facilitates
parallel and independent query evaluation at each tile renders it
a promising approach for distributed spatial data management.
In addition, we plan to investigate the efficiency of our partitioning
scheme on 3D data.
Another direction is to investigate the effectiveness of our secondary
partitioning scheme on other space-partitioning approaches such as the
quad-tree and the kd-tree and the better management of data skew.
Finally, we will study the evaluation of
other popular query types, such as nearest neighbor queries and
spatial joins. 
%Ideas for future work:
%\begin{itemize}
%\item Extend our two-level partitioning to handle distance relationships
%\item Scaling up and out (distributed data management and query evaluation)
%\item Experiments with 3D data
%\item Apply 2-level partitioning on other space-oriented indexes such
% as quad-trees.
%\item nearest neighbor queries and spatial joins
%\end{itemize}

\bibliographystyle{ACM-Reference-Format}
\bibliography{sjoin} 
\end{document}